\documentclass[aps,twocolumn,floatfix,graphics,tightenlines]{revtex4-2}
\usepackage{graphics}
\usepackage{subcaption}
\usepackage{epstopdf}
\usepackage{epsfig}
\usepackage{graphicx}
\usepackage[font=small,labelfont=bf,
   justification=justified,
   format=plain]{caption}
\usepackage{epsf,epic}
\usepackage{array}
\usepackage{color}
\usepackage{amsmath}
\usepackage{booktabs}
\usepackage{multirow}
\usepackage{physics}
\usepackage[hidelinks]{hyperref}
 \usepackage{amsfonts}
\usepackage{wrapfig}
\usepackage{breqn}
\usepackage{pstricks}
\usepackage[utf8x]{inputenc}
\usepackage{multirow}
\usepackage{fancyref}
\usepackage{pst-node}
\usepackage{float}
\usepackage{bm}
\usepackage{dcolumn}
\usepackage{diagbox}

\newcommand{\etal}{\textit{et al.\ }}

\def\dprime{{\prime\prime}}

\makeatletter
\usepackage{etoolbox} 
\appto{\appendix}{%
	\@ifstar{\def\theequation@prefix{A.}}%
	{}%
}
\preto\maketitle{%
  \begingroup\lccode`~=`,
  \lowercase{\endgroup
  \let\saved@breqn@active@comma~
  \let~}\active@comma 
}
\appto\maketitle{%
  \begingroup\lccode`~=`,
  \lowercase{\endgroup
  \let~}\saved@breqn@active@comma 
}
\makeatother
\begin{document}
\title{Band structure and excitonic properties of WSe$_2$ in the isolated
  monolayer limit in an all-electron approach}
\author{Niloufar Dadkhah and Walter R. L. Lambrecht}
\affiliation{Department of Physics, Case Western Reserve University, 10900 Euclid Avenue, Cleveland, OH-44106-7079}
\begin{abstract}
A study is presented of the electronic band structure and optical absorption spectrum of monolayer WSe$_2$ using an all-electron quasiparticle self-consistent $GW$ approach, QS$G\hat W$,   in which the screened Coulomb interaction $\hat W$ is calculated including ladder diagrams representing electron-hole interaction. The Bethe-Salpeter Equation is used to calculate both the screened Coulomb interaction $\hat W$ in the quasiparticle band structure and the imaginary part of the macroscopic dielectric function. The convergence of the quasiparticle band gap and lowest exciton peak position is studied as function of the separation of the monolayers when using periodic boundary conditions. The quasiparticle gap is found to vary as $1/d$ with $d$ the size of the vacuum separation, while the excitonic lowest peak reaches convergence much faster. The nature of the exciton spectrum is analyzed and shows several excitonic peaks below the quasiparticle gap when a sufficient number of \textbf{k} points is used. They are found to be in good agreement with prior work and experiment after adding spin-orbit coupling corrections and can be explained in the context of the Wannier-Mott theory adapted to 2D. 
\end{abstract}
\maketitle
\section{Introduction}\label{sec:intro}
It is well known that 2D materials show strong excitonic effects thanks to the reduced screening in two dimensions \cite{Keldysh79,Cudazzo11} and the intrinsic differences between the 2D and 3D Coulomb problem \cite{Shinada66}. The lower dimensionality also has a strong effect on the quasiparticle self-energy \cite{Delerue03,Komsa12}. Transition metal dichalcogenides are a well studied family of 2D materials which can be isolated in monolayer form. In reality these monolayers are always placed on a substrate and often also capped with additional layers such as hBN. This provides additional screening of the surroundings of the monolayer. Determining the true isolated monolayer properties of a freestanding layer experimentally is thus a challenging task. On the other hand, computationally, determining the isolated limit is also challenging because almost all calculation rely on periodic boundary conditions and thus the convergence as function of interlayer spacing needs to be studied carefully. In this paper, we pick WSe$_2$ as an example to study this question.
\par
Many-body-perturbation theory (MBPT) provides currently the most accurate approach to the electronic band structure in the form of Hedin's $GW$ approach\cite{Hedin65,Hedin69} where $G$ is the one-electron Green's function and $W$ the screened Coulomb interaction. Likewise the Bethe Salpeter Equation (BSE) approach \cite{Strinati88,Hanke78,HankeSham80,Albrecht98,Rohlfing98,Benedict98,Onida02} provides an accurate approach for optical absorption including local field and electron-hole interaction effects. Most of the implementations of these methods start from density functional theory (DFT) in the local density approximation (LDA) or generalized gradient approximation (GGA) and make use of pseudopotentials and plane wave basis sets. Here we use an all-electron implementation based on the full-potential linearized muffin-tin orbital (FP-LMTO) method \cite{Kotani07,questaalpaper}, which uses an auxiliary basis set consisting of interstitial plane waves and products of partial waves \cite{Aryasetiawan94} inside the spheres to represent two-point quantities, such as the bare and screened Coulomb potentials, $v,W$, the polarization propagator $P$ and the inverse dielectric response function $\varepsilon^{-1}$.
\par
Different levels of the $GW$ approximation are in use. The most commonly used method is the $G_0W_0$ approximation in which perturbation theory is used to find the correction due to the difference between the $GW$ self-energy operator and the DFT exchange correlation potential used as zeroth order. Self-consistent $GW$ in the sense conceived by Hedin \cite{Hedin65} does not necessarily improve the results because it appears to require one to include vertex corrections and go beyond the $GW$ approximation.
A successful way to make the results of $GW$ independent of starting point is the quasiparticle self-consistent $GW$ approach (QS$GW$)\cite{MvSQSGWprl,Kotani07}. Nonetheless, this method overestimates the quasiparticle gaps because it underestimates screening. Typically, it underestimates
dielectric constants of semiconductors by about 20 \% as illustrated
for example in Fig. 1 in Ref. \onlinecite{Bhandari18}.
To overcome this problem,
several approaches are possible. One can include electron-hole interactions
in the screening of $W$ by means of a time-dependent density functional
approach by including a suitable exchange-correlation kernel \cite{Shishkin07,ChenPasquarello15}. The approach we use here was recently introduced by Cunningham \etal \cite{Cunningham18,Cunningham23,Radha21} and includes ladder diagrams in the screening of $W$ via a BSE approach for  $W({\bf q},\omega=0)$. It can also be viewed as introducing a vertex correction in the Hedin equation for $W$. No vertex corrections are included in the self-energy
$\Sigma$ but this is justified within the QS$GW$ approach by the approximate cancellation in the ${\bf q}\rightarrow 0$, $\omega\rightarrow0$ limits of the $Z$ factors in the vertex $\Gamma\propto 1/Z$ and the coherent part of the  Green's function $G=ZG^0+\tilde G$ \cite{Radha21}. Here $Z=(1-\partial \Sigma/\partial \omega)^{-1}$ is the renormalization factor from the quasiparticlized $G^0$ to the dynamic $G$ Green's functions and  $\tilde G$ is the incoherent part of the Green's function.  

In this paper, we apply this all-electron approach to the 2D material
WSe$_2$ and study the convergence of various aspects of the method,
such as number of bands included in the BSE approach, QS$GW$ without
and with ladder diagrams, which we call QS$G\hat W$,  and
most importantly the dependence on the size of the vacuum region.
We compare with previous results and with experimental work
and discuss the nature of the excitons.  Our results
confirm the main conclusion of Komsa \etal \cite{Komsa12}
in a study of MoS$_2$ that the long-range effects of the self-energy
in $GW$ lead to a slow $1/d$ convergence of the quasiparticle gap
with $d$ the size of the vacuum region but with a similar and compensating
slow convergence of the exciton binding energy, resulting in a much
faster convergence of the optical gap corresponding to the lowest allowed exciton peak.  
Besides the behavior as function of distance, we show that the convergence of the excitons as function of  the {\bf k} mesh included in the BSE calculations is crucial to obtain good agreement with the exciton gap. We study the spectrum of exciton levels below the quasiparticle gap, including dark excitons and explain their  localization in {\bf k} space and real space in relation to the 2D Wannier-Mott theory. Spin-orbit coupling effects are included in the bands and added as {\sl a-posteriori} corrections to the excitons.

\section{Computational method} \label{sec:method}
The band structure calculations carried out here use the {\sc Questaal} package, which implements DFT and MBPT using a linearized muffin-tin-orbital (LMTO) basis set \cite{questaalpaper}. The details of the QS$GW$ approach and its justification compared to fully self-consistent $GW$ in the Hedin set of equations, can be found in Ref. \onlinecite{Kotani07}. The recent implementation of the BSE in this code is documented in Refs \onlinecite{Cunningham18,Cunningham23,Radha21}. Here we summarize our convergence parameters for its application to WSe$_2$. We use a double ($\kappa$,$R_{sm}$) basis set $spdfgspdf$ on W, with $5p$ semicore-states and high-lying 6d states treated as local orbitals. Here  $\kappa^2$ is the Hankel function kinetic energy and $R_{sm}$ is the smoothing radius of the smoothed Hankel function envelope function.  For Se, we use a $spdfspd$ basis set. The quasiparticle self-energy matrix is calculated up to a cut-off of 2.5 Ry and is approximated by an average diagonal value above 2.0 Ry. Convergence is studied as function of the ${\bf k}$ mesh on which the self energy is calculated and subsequently interpolated to a finer mesh or along the symmetry lines by Fourier transforming back and forth to the real space LMTO basis set. We find that a $9\times9\times3$ set is converged to within 0.1 eV.  Somewhat surprisingly, even for a vacuum region  as large as 35 \AA\  we need more than one ${\bf k}$ division in the short  direction of the reciprocal unit cell perpendicular to the layers in order to obtain a clear $1/d$ behavior of the quasiparticle gap.  This reflects the long-range nature of the self-energy, as will be discussed in more detail later. For the BSE calculations, and the nature of the excitons, on the other hand, a finer in-plane mesh is important. The in-plane lattice constant is chosen as $a=3.32$ \AA\ in the $P\bar{6}m2$ space group, which corresponds to an AA stacking of the trigonal prism unit cell in each layer. The stacking is irrelevant since we focus on the non-interacting monolayer limit. The $c$ lattice constant is varied to study the convergence. The in-plane lattice constant is taken from Materials Project \cite{MaterialsProject,MPwse2} and was optimized in the generalized gradient approximation (GGA). The convergence with $d$, the distance between the layers, is a major object of our study and discussed in Sec. \ref{sec:results}

\section{Results}\label{sec:results}
\subsection{Quasiparticle bands and band gaps}\label{sec:qpgaps}
\begin{figure}[!htb]
    \centering
    \includegraphics[width=1.\linewidth]{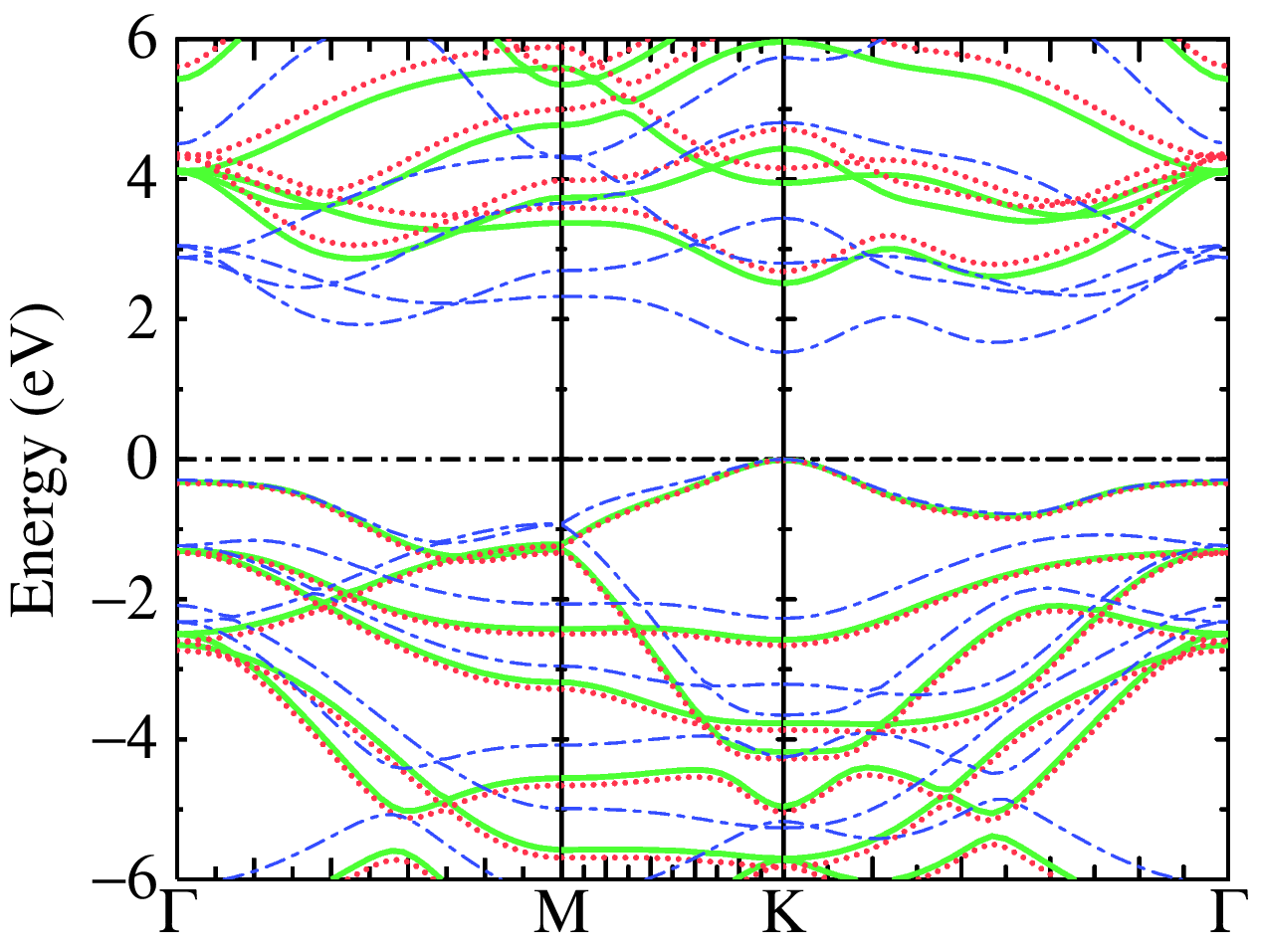}
    \caption{Band structure of WSe$_2$ in the LDA (blue dashed), QS$GW$ (red dotted), and QS$G\hat{W}$ (green solid). The bands are referred to the valence band maximum within each method.}
    \label{fig:bands}
\end{figure}
First, we explore the electronic structure at different levels of theory (DFT, QS$GW$ and QS$G\hat{W}$) for several interlayer distances $d$.  The bandgap is found to be direct with the VBM (valence band maximum) and CBM (conduction band minimum) at points $K$ and $-K$. Fig. \ref{fig:bands} shows the DFT, QS$GW$ and QS$G\hat{W}$ band structure for $d=25$ \AA~specified by blue, red, green, respectively. The band gap changes from 1.53 eV at LDA to 2.95 eV and 2.71 eV in QS$GW$ and QS$G\hat{W}$, respectively using the converged $9\times9\times3$ mesh.  In QS$G\hat{W}$, including the electron-hole coupling in terms of  the ladder diagrams in the calculation of the polarization propagator $P$   reduces $W=(1-Pv)^{-1}v$ to $\hat W$ and hence the self-energy $\Sigma=iG\hat W$ and the gap are also reduced. We note that $(E_g^{\text{QS}G\hat{W}}-E_g^{\text{LDA}})/(E_g^{\text{QS}GW}-E_g^{\text{LDA}})\simeq 0.83$, which shows that the reduction of the gap is in good agreement with the empirical approach of hybrid QS$GW$ (h-QS$GW$), which proposes to  reduce the QS$GW$ self-energy by a universal factor of $\sim$0.8. This approach was found to  lead to good agreement with experimental band gaps for several materials \cite{Deguchi_2016,Bhandari18}. 
\begin{figure}[htbp]
    \centering
        \includegraphics[width=\linewidth,trim=1.3cm 1.3cm 2.5cm 2.5cm, clip]{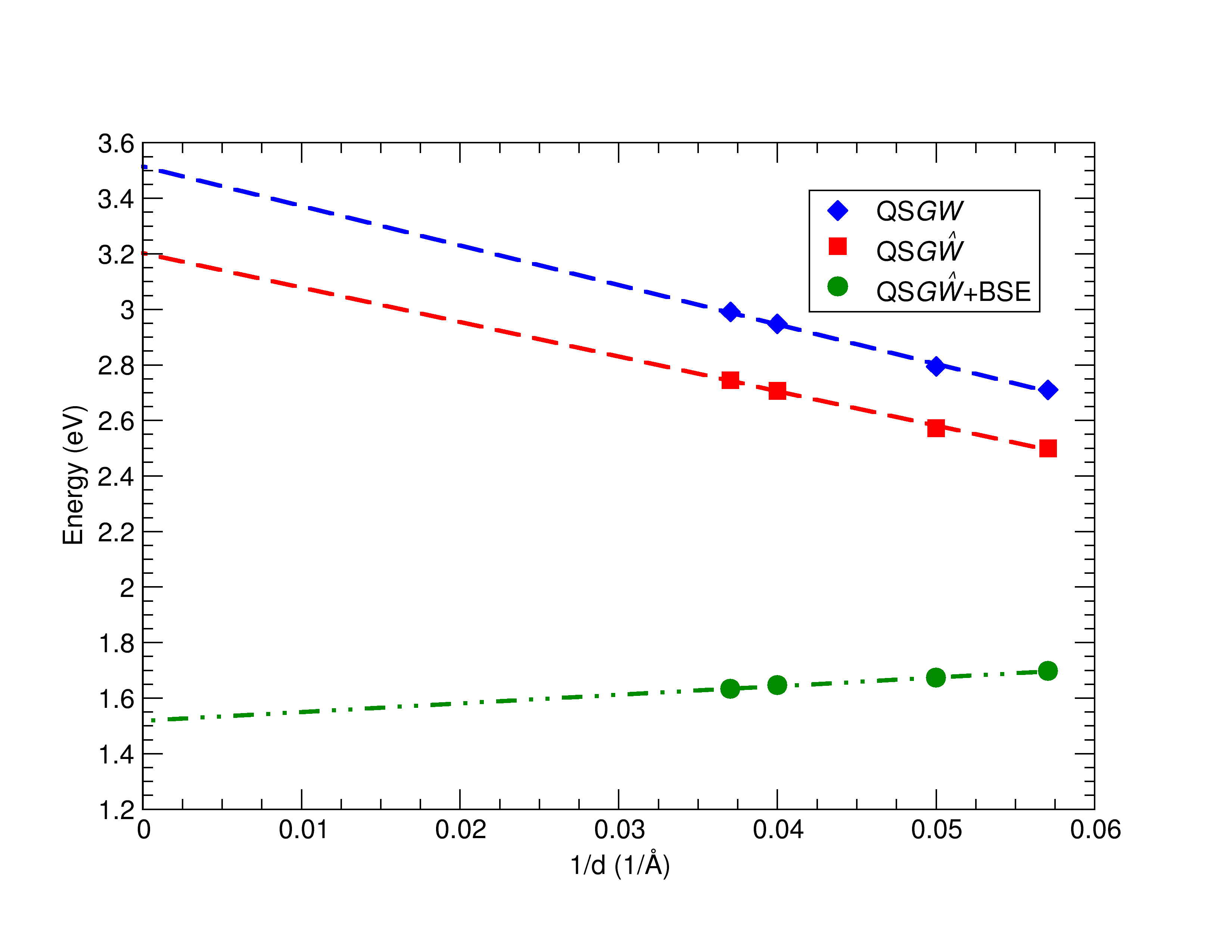}
    \caption{Band gaps in QS$GW$, QS$G\hat{W}$ and QS$G\hat{W}$+BSE (optical gap) as a function of the inverse interlayer distance using a $9\times9\times3$ \textbf{k} mesh at $d=25$ \AA. Dashed lines show the linear fitting to estimate the band gap at $1/d=0$}
    \label{fig:gaps}
\end{figure}

To attain the strict monolayer limit, we change the interlayer distance from 17.5 \AA~up to 27 \AA~and examine the band gaps using different $\mathbf{k}$-mesh samplings. Our results show that it is only for $9\times9\times3$ and $8\times8\times3$ \textbf{k} mesh that we can see a clear linear trend for the gap as a function of $1/d$. One expects a $1/d$ slow convergence because of the long-range nature of the screened Coulomb interaction \cite{Komsa12}. Fig. \ref{fig:gaps} shows the band gaps versus $1/d$ for the $9\times9\times3$ \textbf{k} mesh. We note that although the cell is quite large in the direction perpendicular to the plane, the long-range nature of the self energy requires us to take more than one \textbf{k} point in that direction to achieve convergence. As shown in Fig. \ref{fig:gaps}, QS$GW$ and QS$G\hat{W}$ band gaps converge quite slowly with $d$, such that in the limit for infinite $d$, the extrapolated band gaps are around 3.5 eV and 3.2 eV, respectively which are about 0.7 eV larger than the corresponding gaps at $d=17.5$ \AA. 
\subsection{Optical response}\label{sec:optresponse}
\begin{figure}[htbp]
    \centering
        \hspace{0cm}\includegraphics[width=1.\linewidth,trim=0cm 0cm 0cm 0cm, clip]{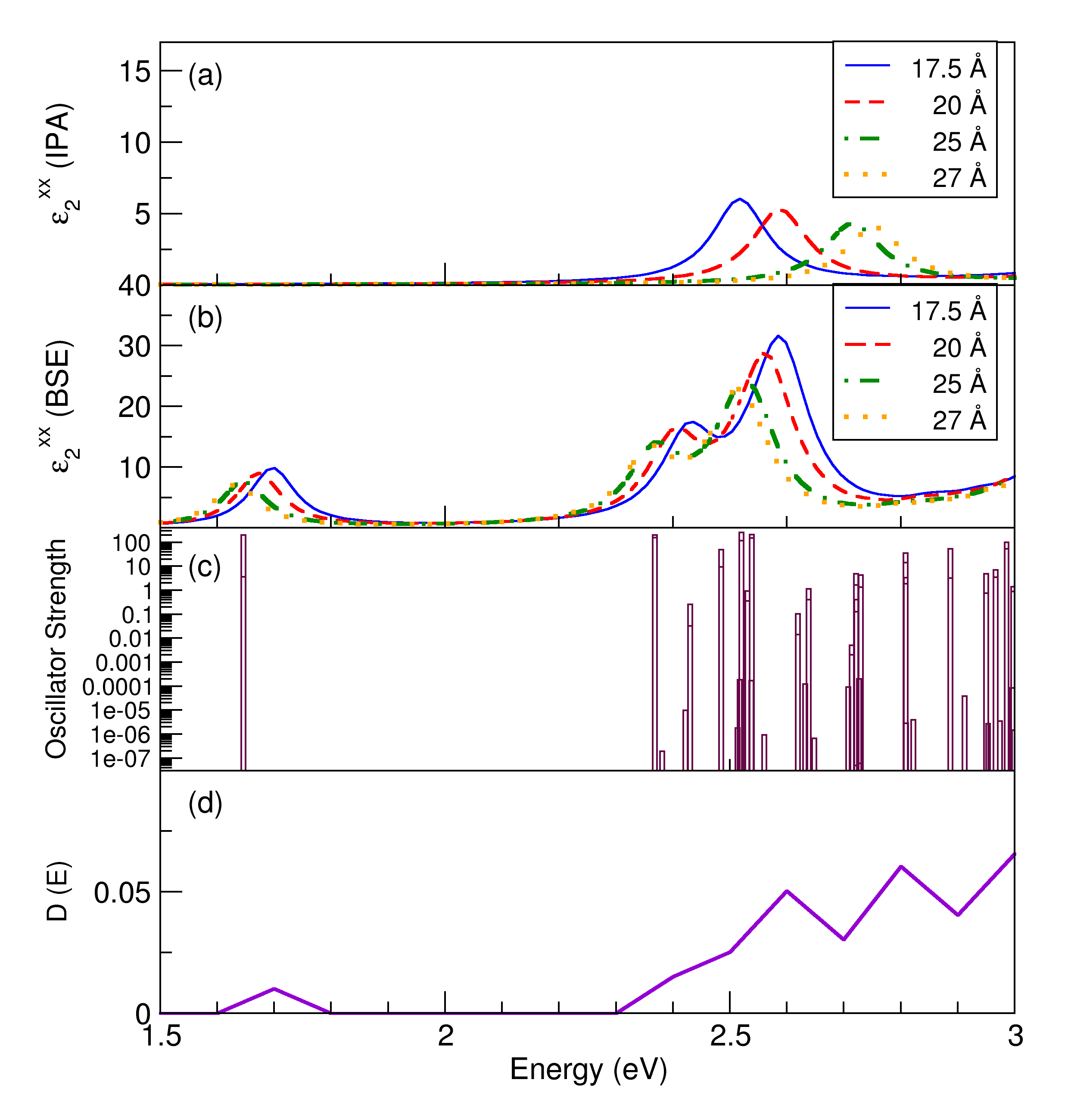}
    \caption{(a)-(b) Imaginary part of the macroscopic dielectric function tensor $\epsilon_2(\omega)$ for $x$ polarization within the IPA  and BSE, respectively; shown for four interlayer distances; (c) Oscillator Strengths along $x$ associated with the excitonic states from the BSE on a logarithmic scale for interlayer distance 27 \AA; (d) Number of excitonic states (on intervals of 0.1 eV) normalized to 1.}
    \label{fig:combined}
\end{figure}
In this section, we turn to the optical absorption, which is closely related to the imaginary part of the dielectric function $\varepsilon_2(\omega)$.  Figure \ref{fig:combined}(a,b) show $\varepsilon_2(\omega)$  for several interlayer distances  as calculated in the Independent Particle Approximation (IPA) and in the Bethe-Salpeter Equation (BSE). The former omits both local field and electron-hole interaction effects and corresponds to vertical transitions between the bands. These calculations have been done using the QS$G\hat{W}$ bands as input and the photon polarization was chosen along the $x$-direction. By symmetry the macroscopic dielectric function is isotropic in the $xy$-plane and we are primarily interested in the in-plane polarization. For the BSE calculations shown in \ref{fig:combined}(b), we include 6 valence and 6 conduction bands in the two-particle Hamiltonian on a $9\times9\times3$ \textbf{k} mesh. We use the Tamm-Dancoff approximation which has been shown to be valid in most  materials \cite{Sander2015}. The spectrum is broadened by including an imaginary part which linearly increases with energy. Although there is an overall red-shift of the spectrum, one can see that the difference between the IPA and BSE does not amount to a simple  shift toward lower energies but BSE significantly modifies the shape of the spectrum. Moreover, in BSE a sharp peak occurs below the quasiparticle gap, which corresponds to the first bright exciton. The position of the various peaks in $\varepsilon_2(\omega)$  does not exhibit significant variation with respect to the interlayer distance. In particular, unlike the QS$GW$ gaps which change strongly with the size of the vacuum region, the position of the first peak or the optical gap stays practically constant for this \textbf{k} mesh. This indicates an increasing binding energy with size of the vacuum region, as expected, given the long range nature and decreasing screening of the screened Coulomb interaction $W$ with size of vacuum region.  This is shown also in Fig. \ref{fig:gaps}, 
which shows that the optical gap is nearly constant as function of interlayer distance within the range of distances shown. In fact, it appears to slightly decrease with layer distance, but this variation is within the error bar of the calculations. It may be due to a slightly decreasing convergence of the basis set for larger interlayer distances. Furthermore, as we discuss in the next section, the exciton peak is sensitive to the {\bf k}-mesh sampling. In the result shown in Fig. \ref{fig:gaps} the mesh is kept fixed at the same mesh as used in the QS$G\hat{W}$ band calculation, and provides only a lower limit of the optical or exciton gap. 

Fig. \ref{fig:combined}(c) shows  the oscillator strengths for all eigenvalues of the two-particle Hamiltonian on a log-scale and Fig. \ref{fig:combined}(d) shows the density  of excitonic states (normalized to 1), both at $d=27$ \AA. The difference between the density of excitonic states and the actual $\varepsilon_2(\omega)$ indicates the modulation of the spectrum by the dipole matrix elements. Fig. \ref{fig:combined}(c) shows that the matrix elements can vary by several orders of magnitude. Only the levels with highest oscillator strength  in (c) show up as peaks in (b). The oscillator strengths here are not normalized and thus only the relative intensity between different levels has physical meaning.
\subsection{Convergence of exciton gap}\label{sec:excitongap}
\begin{figure} 
    \centering
        \includegraphics[width=\linewidth,trim=1cm 1cm 5.5cm 6.5cm, clip]{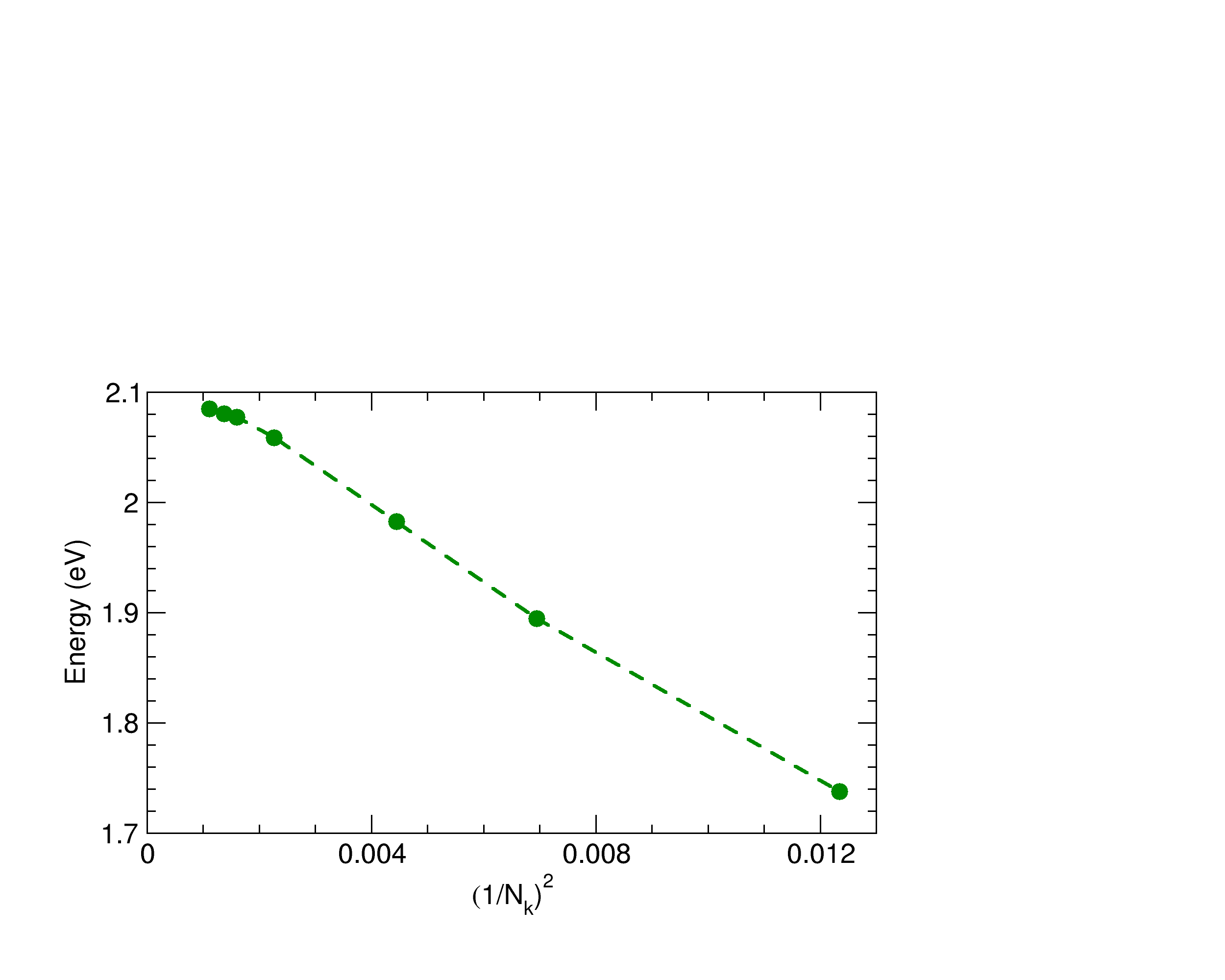}
    \caption{Energy of the lowest exciton as a function of the inverse of the number of {\bf k} points squared for $d=20$ \AA.}
    \label{fig:firstex-vs-nkgw}
\end{figure}

To converge the energy of the first exciton peak, it is necessary to use a finer {\bf k} mesh, since it is localized in {\bf k} space. On the other hand, as we will show in \ref{sec:exanalysis}, for the lowest excitons, we can restrict the active space of bands to fewer valence and conduction bands included in the BSE. Fig. \ref{fig:firstex-vs-nkgw} shows the exciton gaps as function of $1/N_{k}^2$, where $N_{k}$ is the total number of {\bf k} in the 2D mesh and for a distance $d=20$ \AA, which is close to the converged monolayer limit and including 4 valence and 4 conduction bands. For the direction normal to the plane, three {\bf k} points are used. In fact, as mentioned earlier, from Fig. \ref{fig:gaps} the BSE calculated lowest exciton peak does not vary significantly with $d$. The optical gap for the isolated monolayer limit and for $N_k\rightarrow\infty$ is thus found to be $\sim$2.1 eV, whereas the quasiparticle gap in that limit was found to be 3.2 eV, implying an exciton binding energy of 1.1 eV. 
\subsection{Discussion of exciton spectrum in relation to prior theory work and experiment} \label{sec:spectrum}
\begin{figure}
       \includegraphics[width=1.\linewidth,trim=1cm 1.7cm 8.2cm 8.3cm, clip]{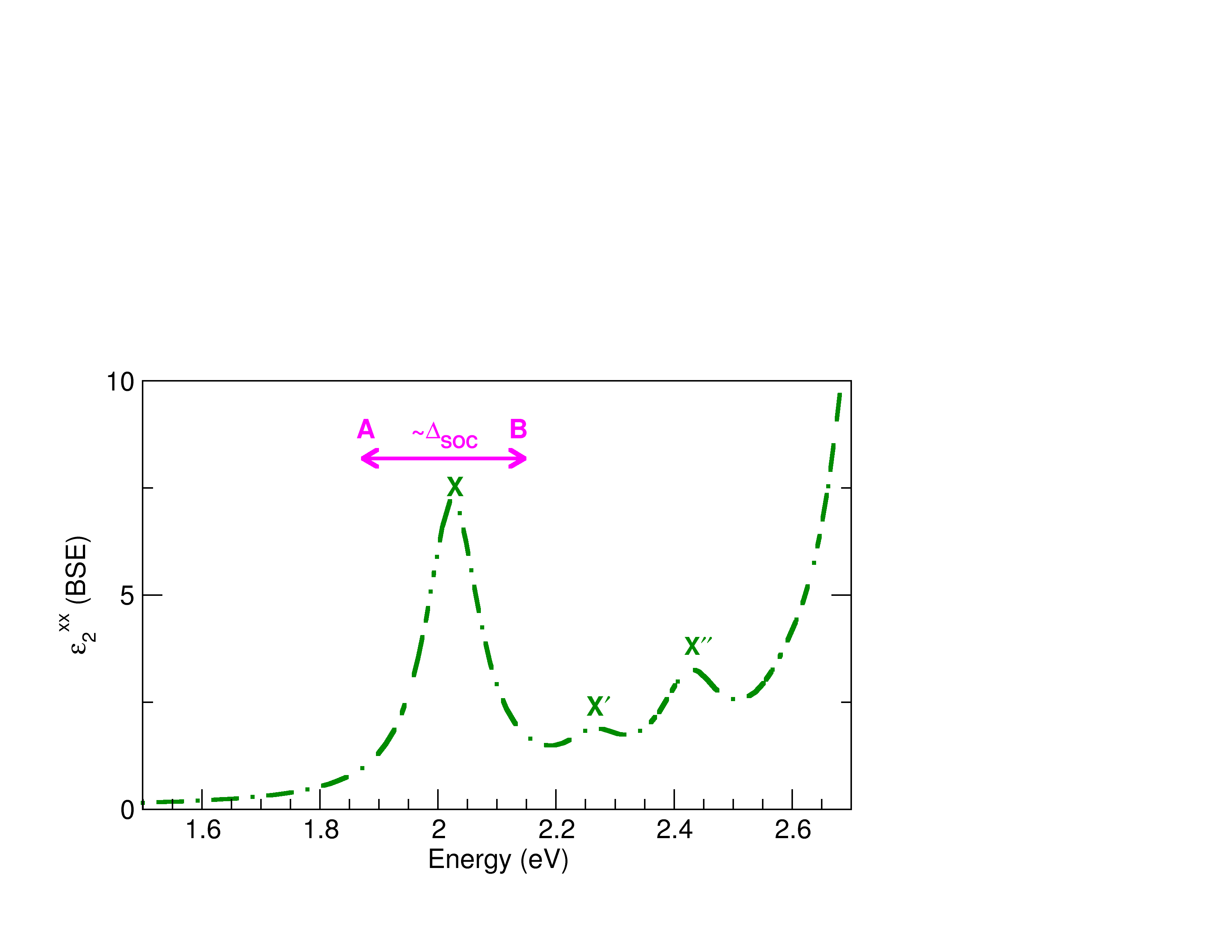}
    \caption{Imaginary part of the macroscopic dielectric function tensor $\epsilon_2(\omega)$ for $x$ polarization within the BSE using one valence and one conduction band on $30\times30\times3$ {\bf k} mesh. Label X shows the ground state exciton without SOC, know as $1s$. Adding SOC, splits X exciton into two excitons commonly known as excitons A and B, which are schematically shown as magenta A and B labels. X$^\prime$ shows the second bright exciton, known as $2s$. A third prominent peak also appears, which is labeled as X$^{\dprime}$. Here, the spectrum is shifted so that the energy of the first peak coincides with the converged value of 2.1 eV.}
    \label{fig:epszoom}
\end{figure}

Fig. \ref{fig:epszoom} shows $\varepsilon_2(\omega)$ in the BSE as in Fig. \ref{fig:combined}(b), for $d=25$ \AA, but using a finer {\bf k} mesh of $30\times30\times3$ with $N_v=N_c=1$.    This finer mesh provides us with a better picture of the profile of the spectrum below and just above the quasiparticle gap. We can readily notice that compared to Fig. \ref{fig:combined}(b), more weight is shifted to the first peak, making it the strongest optical excitation.
Table \ref{tab} shows some of the experimental and theoretical results for the energy of A exciton (first bright), binding energy of the A exciton and the energy splitting between A and A$^\prime$(second bright). To compare our results here with experiment and other literature work, it is important to include spin-orbit coupling (SOC) effects. {\sc Questaal}'s BSE code does not yet allow for the inclusion of SOC; however, we can include spin-orbit coupling perturbatively in our QS$G\hat{W}$ calculations. In other words, we add the SOC Hamiltonian to the QS$G\hat{W}$ Hamiltonian matrix in the LMTO basis set and rediagonalize it. We don't just evaluate its effect on each separate eigenstate but we do not include SOC effects in the calculation of the self-energy and do not use a full spin-dependent Green's function in the $GW$ steps. Adding SOC splits the lowest exciton peak into two commonly known as A and B excitons which results from the SOC splitting of the highest valence and lowest conduction bands at and around point $K$  \cite{Tawinan12}. Our results show that at point $K$, the valence band splitting is about 445 meV with the VBM-B going 225 meV down and VBM-A 219 meV up in energy. However, the CBM, which has a much smaller splitting also shifts down for both spin directions by about 0.035 eV. Including these {\sl a-posteriori} SOC effects, the A exciton in our results is placed at  $\sim$1.8 eV which is in excellent with results from \cite{Marsili2021PRB}. These authors include SOC fully as well as perturbatively and it is with the latter that we best agree. Most of the experimental results also place the A exciton at about 1.7-1.8 eV, whereas some older theoretical calculations \cite{Ramasubramaniam12} find it near 1.5 eV. 
In Fig. \ref{fig:epszoom} we label our exciton peaks as X, X$^{\prime}$ and X$^{\dprime}$ to indicate that they do not include SOC. Each would split into an A and B peak when adding SOC. 

Our calculated binding energy of $\sim$1 eV is slightly higher  than previously reported. Ramasubramaniam \cite{Ramasubramaniam12} obtains a gap of 2.42 eV at K, but his calculation includes spin-orbit coupling (SOC). If we compare with the average of the SOC split states, his quasiparticle gap would be 2.67 eV but this is at a interlayer separation $d=15$ \AA. 
Furthermore, this author obtained a lower indirect $K$-$\Lambda$ gap from the valence band at $K$ to a secondary valley in the conduction band along the $\Gamma$-$K$ axis $\Lambda$. 
In Fig. \ref{fig:gaps} the smallest distance we considered is 17.5 \AA\ and it already gives a slightly higher gap. Our extrapolated quasiparticle gap for infinite distance is significantly higher. Ramasubramaniam's lowest exciton peak occurs at 1.52 eV for the A exciton and 2.00 eV for the B exciton, showing that this splitting simply corresponds to the corresponding valence band SOC splitting at K. His exciton binding energy is thus reported as 0.9 eV. He used a {\bf k} mesh of $6\times6\times1$. 
We can see that our exciton binding energy is only slightly higher than his while our optical gap and quasiparticle gap are  both significantly higher but in better agreement with \cite{Marsili2021PRB} and experiment. 

Moreover, Fig. \ref{fig:epszoom} shows that between the lowest exciton and the quasiparticle gap, other peaks occur and by examining the oscillator strengths, we see that there are also several dark excitons in this region as shown in Fig. \ref{fig:combined}(c). We identify the next bright exciton which is slightly lower in intensity but clearly visible in the $\varepsilon_2(\omega)$ in Fig. \ref{fig:epszoom}  as the X$^\prime$ exciton, which in literature is usually associated with either a $2s$ or $2p$ type exciton using the hydrogenic Rydberg series nomenclature. Whether $2s$ or $2p$ are showing up in experiment depends on the experimental probe used: two-photon absorption reveals the $2p$ excitons, while linear absorption or luminescence reveals the one-photon absorption. This arises from the different matrix elements and symmetry selection rules. We should note that this classification does not imply a strict hydrogenic Rydberg series, because the system is 2D rather than 3D and the screened Coulomb potential does not behave simply as $1/\varepsilon r$ because the screening in 2D is strongly distance-dependent \cite{Qiu13}. Furthermore, a $2p$ type exciton having an envelope function which has nodes would be dark in the $\varepsilon_2(\omega)$ spectrum but accessible in two-photon absorption experiments. Adding SOC will split this peak into two peaks known as A$^{\prime}$ and B$^\prime$. Following our previous discussion of including SOC effects, We expect the splitting between A and A$^{\prime}$ to stay almost the same as X and X$^{\prime}$ since we associate them to $s$-like states. We call the next peak X$^{\dprime}$, which is close to X$^{\prime}$ in energy but relatively higher in intensity. We also caution that the details of the exciton spectrum shown in Fig. \ref{fig:epszoom}, such as peak splittings and positions depend strongly on the convergence with {\bf k} points. However, qualitatively Fig. \ref{fig:epszoom} is in good agreement with the results of \cite{Marsili2021PRB} without SOC and shows the same relative intensities for X, X$^{\prime}$ and X$^{\dprime}$. In particular, our X-X$^\prime$ splitting is about 255 meV which is very close to the reported value of 215 meV by \cite{Marsili2021PRB} for a $39\times39\times1$ {\bf k}-mesh sampling. This value is higher than values in Table \ref{tab}, which the authors contribute to the substrate effects used in experiments. These authors also report a value of 140 meV using a coarser {\bf k} mesh, which is closer to experimental results.

We conclude that both convergence as function of {\bf k} mesh and vacuum thickness plays an important role. This was also concluded by Qiu \etal \cite{Qiu16} in the case of MoS$_2$. These authors in fact dealt with the monolayer separation by using a truncated Coulomb potential and used an even finer {\bf k} mesh. Their use of a truncated Coulomb interaction allows for an easier convergence to isolated monolayers and they compare the screening dielectric constant with the strict 2D Keldysh limit \cite{Keldysh79}. 
\begin{table*}
  \caption{A exciton energy, A exciton binding energy and A-A$^\prime$ energy splitting of monolayer WSe$_2$ as reported in several experimental and theoretical studies.}
  \begin{ruledtabular}
    \begin{tabular}{cccc}
       & Method & Optical gap (eV) /E$_b$ (meV) & A-A$^\prime$ splitting (meV) \\ [0.5ex] \cline{2-4} \\[-2ex] 
      hBN/ML-WSe$_2$/hBN \cite{Chen2019NanoLett} & Magneto-PL & 1.727/170 & 131 \\
      hBN/ML-WSe$_2$/hBN \cite{Liu2019PRB} & Magneto-PL & 1.712/172 & 131 \\
      ML-WSe$_2$/Si/SiO$_2$ \cite{He2014PRL} & Linear absorption and two-photon PL & 1.65/370 & 160 \\
      hBN/ML-WSe$_2$/hBN \cite{Stier2018PRL}  & Magnetoabsorption & 1.723/161 & 130 \\
      hBN/ML-WSe$_2$/hBN \cite{Liu2021NatCommun} & Gate-dependent reﬂection and PL & $\sim$1.72/- & 120 \\
      hBN/ML-WSe$_2$/hBN \cite{Wang2020NanoLett} & Magneto-photocurrent & 1.725/168.6 & 128.6 \\
      hBN/ML-WSe$_2$/hBN \cite{Woo2023PRB} & EELS & 1.697/- & - \\
      ML-WSe$_2$/Si/SiO$_2$ \cite{Arora2015Nanoscale} & PL and reﬂectance & 1.744/- & - \\
      ML-WSe$_2$/PC \cite{Schmidt20162DMater} & Absorption & $\sim$1.66/710 & 200 \\
      ML-WSe$_2$ \cite{Hong2020PRL} & Momentum-resolved EELS & 1.69/650 & - \\
      ML-WSe$_2$/Si/SiO$_2$ \cite{Wang2015PRL} & 1- and 2-photon PL excitation & 1.75/600$\pm$200 & 150 \\
      ML-WSe$_2$/SiO$_2$ \cite{Li2014PRB} & Reflectance & 1.67/- & - \\
      hBN/ML-WSe$_2$/hBN \cite{Wierzbowski2017SciRep} & PL & $\sim$1.75/- & - \\
      ML-WSe$_2$/hBN/n-doped Si \cite{Madeo2020Sicence} & Photoemission & 1.73/390$\pm$150 & - \\
      hBN/ML-WSe$_2$/hBN \cite{Li2020PRL} & Magnetoabsorption & 1.722/- & - \\
      ML-WSe$_2$\cite{Ramasubramaniam12} & $GW$-BSE & 1.52/900 & - \\
      ML-WSe$_2$\cite{Marsili2021PRB} & $GW$-BSE & $\sim$2.1/596 & - \\
      ML-WSe$_2$\cite{Marsili2021PRB} & $GW$-BSE-Pert. SOC & $\sim$1.8/533 & 140 \\
      ML-WSe$_2$ \cite{Marsili2021PRB} & $GW$-BSE-SOC & $\sim$1.7/550 & 140 \\
      ML-WSe$_2$ \cite{Deilmann2017PRB} & LDA+$GdW$+BSE+SOC & 1.76/- & - \\
      ML-WSe$_2$ \cite{Ramasubramaniam12} & $G_0W_0$+BSE+SOC & 1.52 \\
       ML-WSe$_2$ \cite{Wang2015PRL} & $GW_0$+BSE & $\sim$2 \\
      ML-WSe$_2$ \cite{Deilmann20192DMater} & $GW$+BSE & 1.78/620 & -
    \end{tabular}
  \end{ruledtabular}
  \label{tab}
\end{table*}
\subsection{Exciton wave function analysis}\label{sec:exanalysis}
\subsubsection{Overview} \label{sec:overview}
To better understand the nature of excitons, we can analyze the \textbf{k}-space and real-space distribution of the first few excitons.  First, we calculate $A_{\lambda}^{v,c}(\mathbf{k})=\braket{v,c,\mathbf{k}}{\Psi_{\lambda}^{Ex}}$ for a particular band pair $\{v,c\}$ along several high-symmetry paths in the BZ where $\lambda$ denotes the excitonic eigenstate. Here we focus on the first few excitons by separately considering the ranges of energies which include only $\lambda=1,2$ (degenerate-bright), $\lambda=3$ (dark), $\lambda=4$ (dark), $\lambda=5,6$ (degenerate-semi-bright), $\lambda=7,8$ (degenerate-semi-bright) and $\lambda=21,22$ (degenerate-bright). We should note that bright excitons $\lambda=1,2$, $\lambda=5,6$ and $\lambda=7,8$ correspond to X, X$^\prime$ and X$^\dprime$ peaks in Fig. \ref{fig:epszoom}, respectively. 

In this section we use $N_v=1$, $N_c=1$, which then allows us to use a finer {\bf k} mesh. This will be justified  in Sec. \ref{sec:bandweights}. However, we have checked that they give essentially the same results 
for the exciton region below the quasiparticle gap as a $N_v=4$, $N_c=4$
for the smaller {\bf k} meshes used.  For a study of the overall $\varepsilon_2(\omega)$ up to higher energies in the continuum, it is important to include more bands but this limits the number of {\bf k} points we can afford. When focusing on the excitons, the opposite is true  and the important convergence parameter
is the number of {\bf k} points. 

Fig. \ref{k} in section \ref{sec:bandweights} shows the band weight of various excitons.  We use the weight $W_\lambda^c({\bf k})=\sum_v|A^{vc}_\lambda({\bf k})|^2$ at each {\bf k} by the size of the colored circle in the conduction bands $c{\bf k}$ and likewise $W_\lambda^v=\sum_c|A^{vc}_\lambda({\bf k})|^2$ for the weight on the valence bands. The colors only distinguish between different bands but have no physical meaning. 

Next, we can investigate the $A_\lambda^{v,c}({\mathbf k})$ for a particular exciton and band pair as function {\bf k}  on a mesh of {\bf k} points in Sec, \ref{sec:kspace}. These exciton expansion coefficients, in fact are the Fourier transforms of a slowly varying envelope function within the Wannier-Mott ${\bf k}\cdot{\bf p}$ theory of excitons \cite{Qiu16}. This slowly varying envelope function would be a hydrogenic function in case of a 3D isotropic system and the full real-space exciton function is then modulated by the product of the Bloch functions at the valence and conduction band edges at the ${\bf k}$ point where the exciton is centered.  Here they are 2D analogs and somewhat distorted by the anisotropies. In 2D the angular behavior of the envelope function is described  by the 2D Laplacian which has solutions of the form $\cos{m\phi}, \sin{m\phi}$ and $m=0,1,2,\dots$ are referred to as $s,p,d,\dots$ in analogy with the 3D case. The general solution $e^{im(\phi-\phi_0)}$ contains an arbitrary phase, resulting from the overall arbitrary phase of the eigenstates of the two-particle Hamiltonian diagonalization, but this only affects the orientation of these exciton wavefunctions in {\bf k} space.  This is valid as long as the slowly varying potential for which the slowly varying envelope function is a solution of the effective Schr\"odinger equation, is indeed axially symmetric. This is true for {\bf k} points close to the point $K$ in the region contributing to the exciton. Beyond a certain distance in {\bf k} space, one might expect a trigonal warping. The real-space excitons at short distance thus will not show a circular but rather hexagonal shape. Conversely, the less strongly bound excited excitons are more spread in real space and hence more localized in {\bf k} space and hence in the region where circular symmetry is valid. The main advantage of this analysis is that it shows us the sign changes of the envelope functions as function of angle and hence the angular patterns, which can explain why certain excitons are dark. 

Finally, we will examine the excitons fully in real space in Sec. \ref{sec:realspace}. In that case, we look at $|\Psi({\bf r}_h,{\bf r}_e)|^2$ and we can either fix the hole at ${\bf r}_h$ or the electron at ${\bf r}_e$ and then plot the probability distribution of the other particle. Since the valence and conduction bands, from which the excitons are derived, are all mainly W-$d$ like,  either choice is equally good and we choose to fix the hole. These will be shown in Sec. \ref{sec:realspace}. These will show both the large scale envelope function behavior and the local atomic orbital character. 

Before we present our exciton analysis results let us mention that our discussion of bright and dark excitons here is limited to singlet excitons. Excitons can also be dark due to the spin structure as was discussed for WSe$_2$ and MoS$_2$ by Deilman \etal \cite{Deilmann2017PRB}. The darkness of excitons considered here is rather related to the excited nature of the excitons in the Rydberg series, and their slowly varying envelope function in a Wannier-Mott exciton model.  

We also note that all bright excitons are doubly degenerate and all dark excitons are singly degenerate. The double degeneracy of the first bright exciton results from the equal contributions at $K$ and  $-K$. These  are related by time reversal symmetry and hence, there is a Kramers degeneracy. Their double degeneracy can also be understood from the symmetry analysis of Song \etal \cite{Song2013}. 
In fact, the exciton is expected to be localized near the W atom since the top valence and bottom conduction bands are both dominated by W orbitals.  The excitons can thus be classified according to the symmetry of the W-site which is the full symmetry $D_{3h}$  of the crystal.  Because we focus on excitons with  polarization in the plane, this immediately tells us that all bright excitons will be doubly degenerate and correspond to the $E'$ representation. Dark excitons which are symmetric with respect to the horizontal mirror plane (as we find to be the case for all the ones we examined) must thus correspond to the $A_1'$ or $A_2'$ irreducible representations which are both non-degenerate. The distinction between the two is whether they are even or odd with respect to the vertical mirror planes and or 2-fold rotations in the plane.

\subsubsection{Bandweights} \label{sec:bandweights}
\begin{figure*}
\includegraphics[width=0.45\textwidth,trim=1.3cm .5cm 1.3cm .45cm, clip]{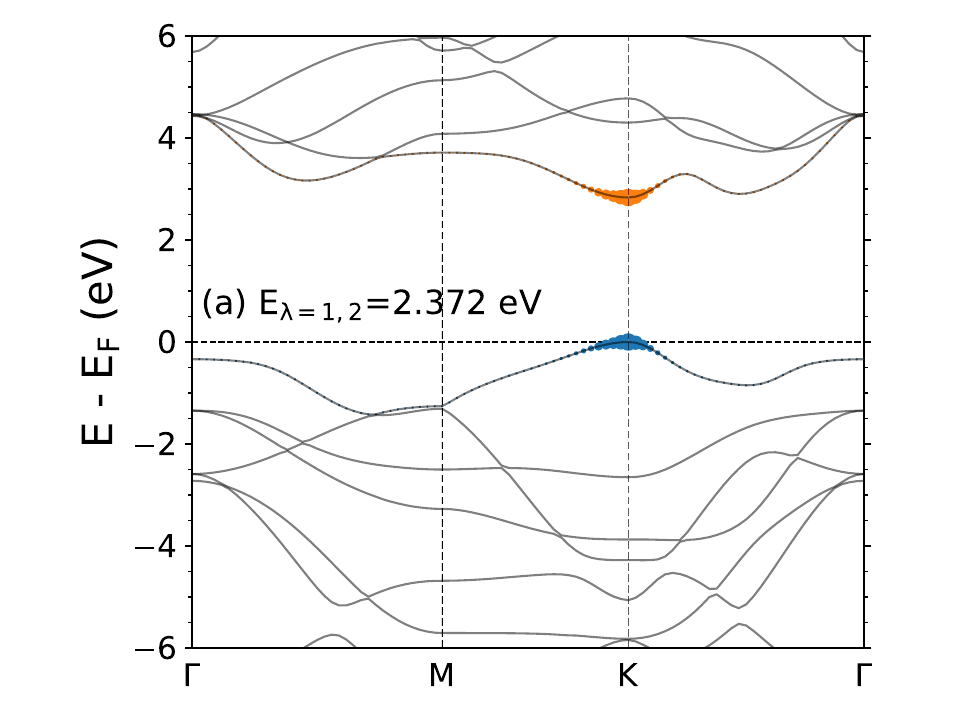}
\includegraphics[width=0.45\textwidth,trim=1.3cm .5cm 1.3cm .45cm, clip]{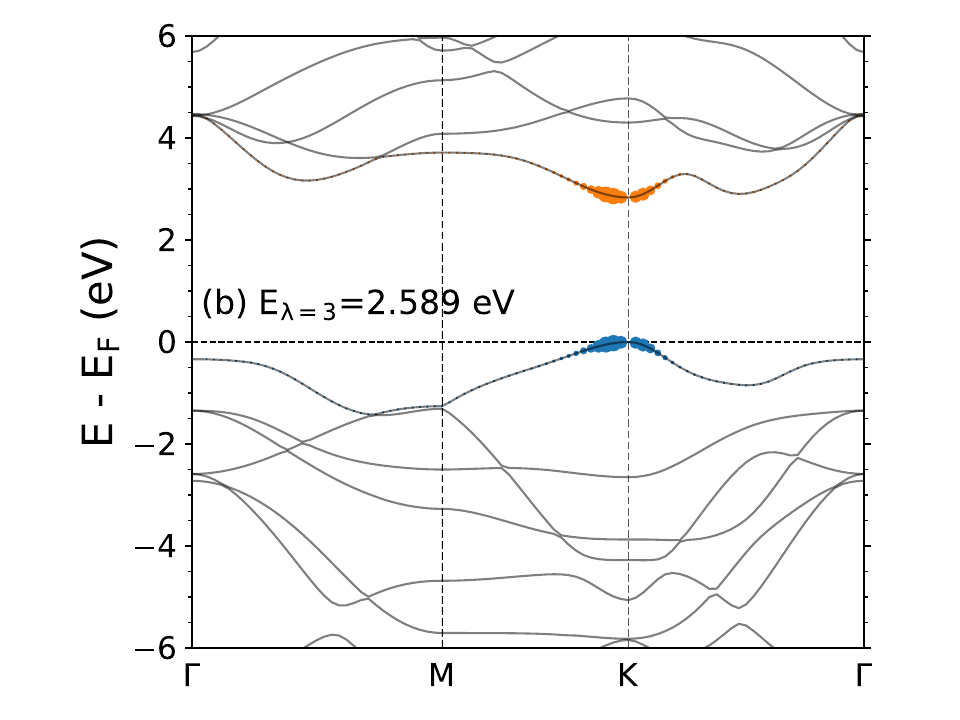}
\includegraphics[width=0.45\textwidth,trim=1.3cm .5cm 1.3cm .45cm, clip]{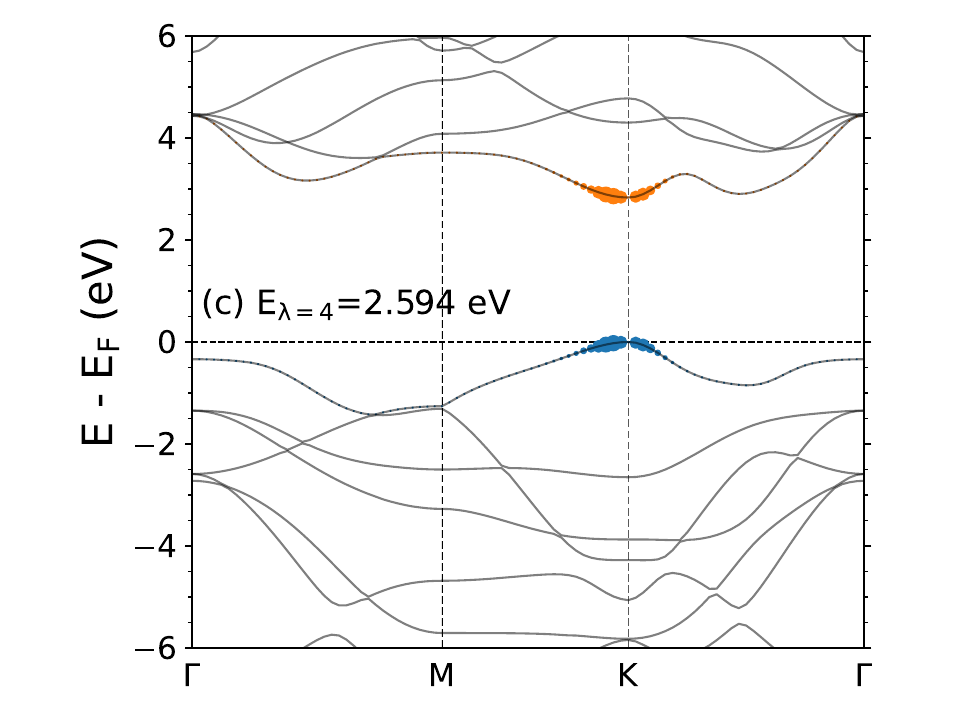}
\includegraphics[width=0.45\textwidth,trim=1.3cm .5cm 1.3cm .45cm, clip]{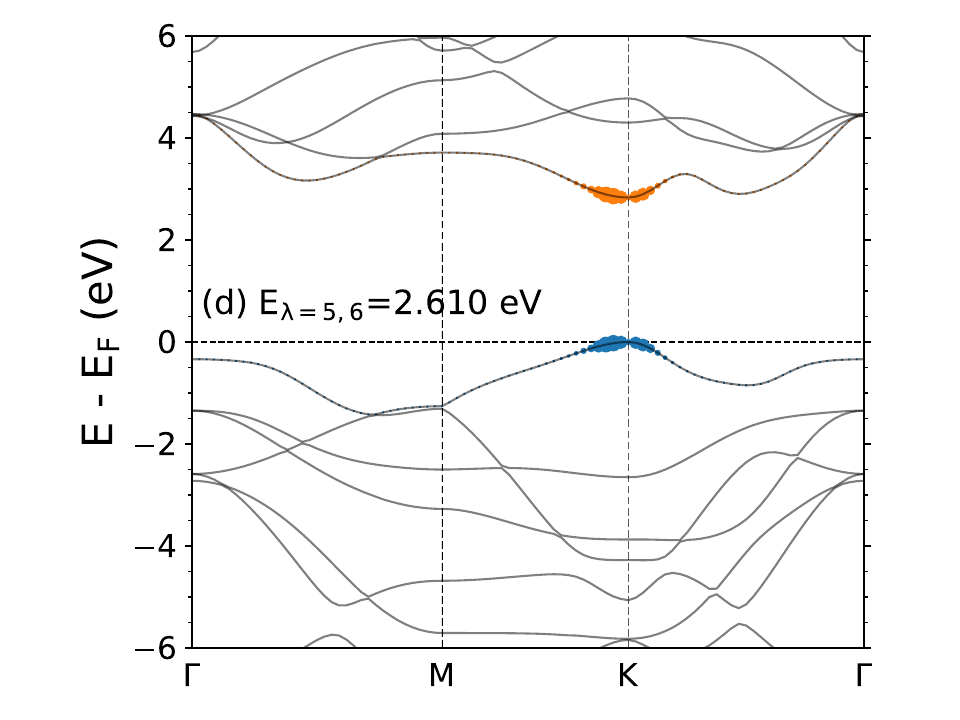}
\includegraphics[width=0.45\textwidth,trim=1.3cm .5cm 1.3cm .45cm, clip]{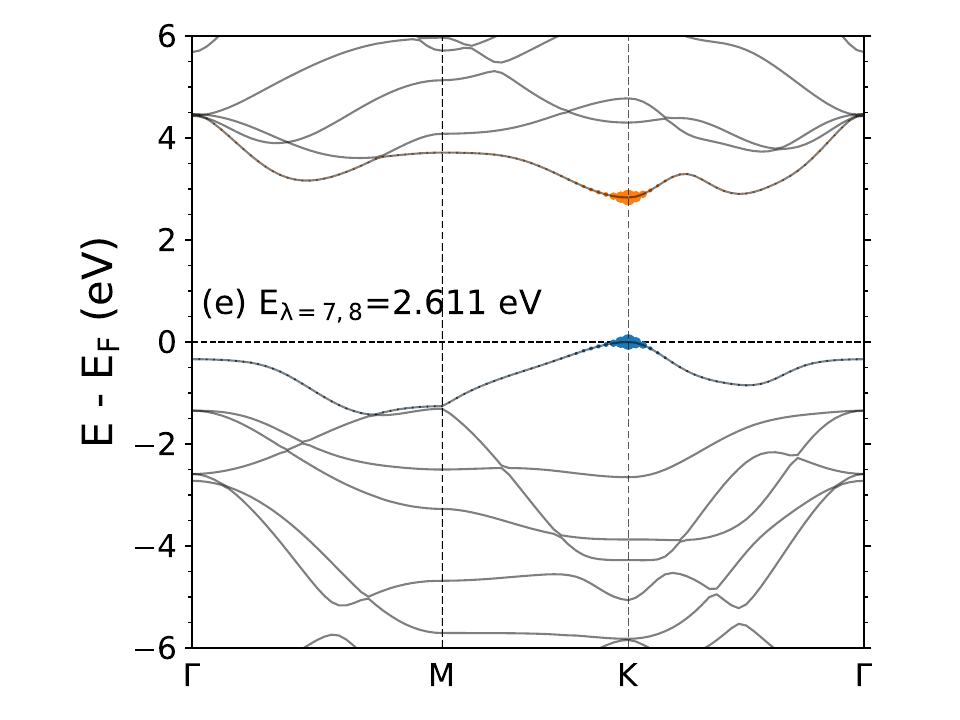}
\includegraphics[width=0.45\textwidth,trim=1.3cm .5cm 1.3cm .45cm, clip]{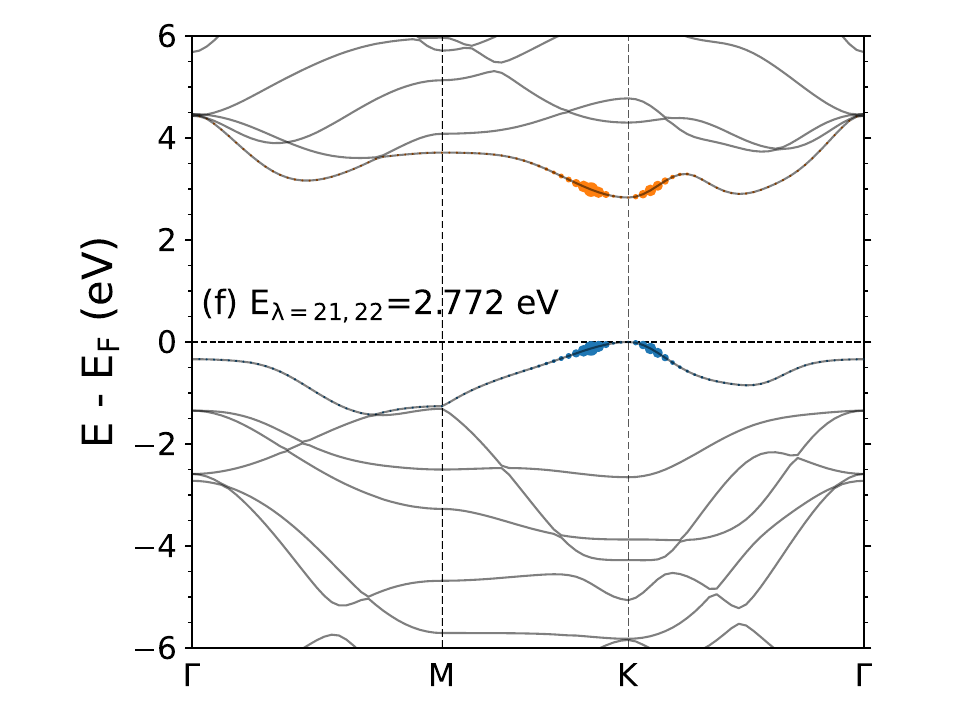}
\caption{Weights of exciton wave function contributed by different bands (a)--(f) for six excitonic eigenvalues. The size of the colored circles indicates the exciton weight $|A^\lambda_{vc{\bf k}}|^2$ for a given exciton $\lambda$. The colors 
    have no meaning and only serve to distinguish different bands.}
    \label{k}
\end{figure*}

Fig. \ref{k} shows the band weights for various excitons. We can clearly see that almost all the contribution comes from the pair of the highest valence and lowest conduction bands focused on and around the point $K$ in the Brillouin zone and decaying as we move away from $K$ in an anisotropic way. For example the conduction band has a larger mass along the $KM$ than the $K\Gamma$ direction and the exciton spreads out further along the $KM$ direction. We can explain this in terms of the Wannier-Mott hydrogenic model of the exciton. First, indeed, the localization in {\bf k} space implies a large spread in real space consistent with the Wannier-Mott type exciton. Second, within this model the effective Bohr radius of the exciton is inversely proportional to the band effective mass and hence gives a smaller Bohr radius in real space for the $MK$ direction which corresponds to a larger spread in {\bf k} space. Figures \ref{k}(b) and \ref{k}(c) demonstrate the weights for the second and third excitons in energy, which are $|A_{\lambda=3}^{vc\mathbf{k}}|$ and $|A_{\lambda=4}^{vc\mathbf{k}}|$, respectively. For these two dark and non-degenerate excitons, we can clearly see that almost all the contribution still comes from only the pair of the highest valence and lowest conduction bands focused around the point $K$ in the Brillouin zone but now with a zero value at point $K$ itself in clear contrast to the first bright exciton where the contribution at point $K$ is the largest, as shown in Fig. \ref{k}(a). Fig. \ref{k}(d), shows the weight $|A_{\lambda=5}^{vc\mathbf{k}}|, |A_{\lambda=6}^{vc\mathbf{k}}|$ which corresponds to fourth exciton in energy, doubly degenerate and semi-bright, {\sl i.e.} less bright than the first bright exciton.  This exciton has characteristics in between those  of the previous bright and dark excitons, with an overall distribution like the dark excitons but nonzero contribution at point $K$. Fig. \ref{k}(e) corresponds to fifth exciton in energy which resembles the first bright exciton in oscillator strength  but less extended around $K$. Fig. \ref{k} (f) shows the excitonic weights for an exciton with higher energy which, in its {\bf k}-space distribution, resembles the dark excitons even though it has a strong oscillator strength. In {\bf k} space we see it has larger dip at and around point $K$.

\subsubsection{{\bf k}-space analysis}\label{sec:kspace}
To explain the behaviors noted in Sec. \ref{sec:bandweights}, we need to fully examine the {\bf k}-space relations of the $A_\lambda^{vc}({\bf k})$ instead of only looking at their absolute values along high-symmetry lines. 
Because the bright excitons are doubly degenerate, any linear combination of the two eigenstates  is also an eigenstate and this makes it  difficult to visualize them. Thus, in this section we focus  on the dark excitons. 

\begin{figure*}
  \includegraphics[width=1\textwidth]{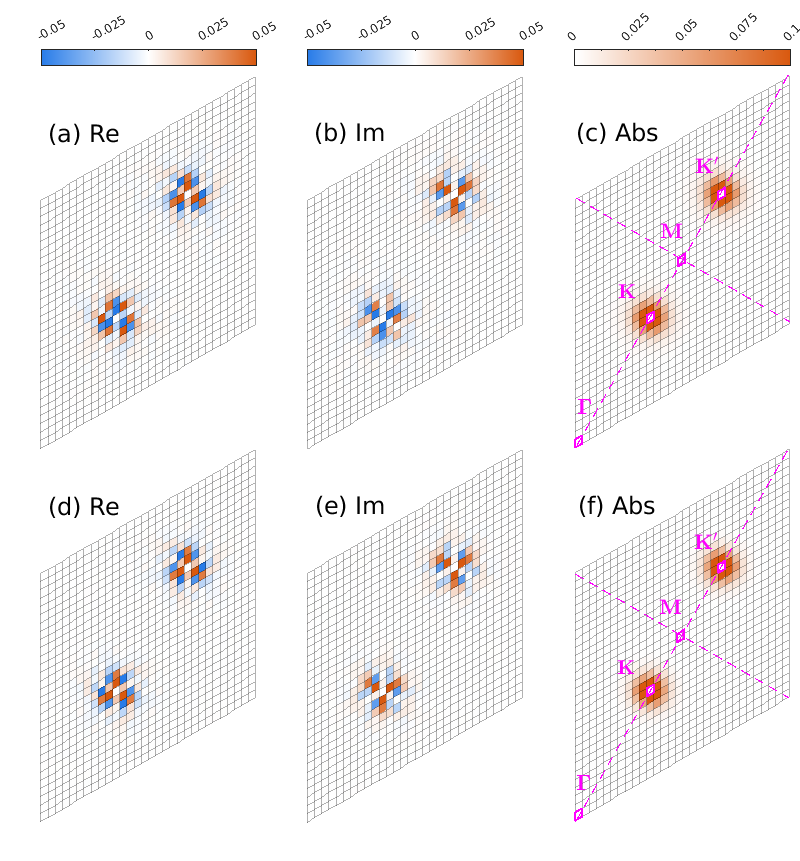}
  \caption{Real, imaginary and absolute values of the envelope function $A^\lambda_{vc{\bf k}}$ for the first two dark excitons over the $\mathbf{k}$ space. First and second rows belong to $\lambda=3$ and $\lambda=4$, respectively. High symmetry points are labeled as magenta in the absolute value panel as reference.}
    \label{kdark}
\end{figure*}

In Fig. \ref{kdark} we show the real, imaginary and absolute values of the envelope function $A_\lambda^{vc}({\bf k})$ for the first two dark excitons over the 2D {\bf k} space in one unit cell as a color plot. This requires an even finer {\bf k} mesh but we can now restrict the $v$ and $c$ to just the valence and conduction band edges. We use here a $30\times30$ mesh. In spite of the rather coarse pixelation, we can see that around both the points $K$ and $K'=-K$ there is a three fold symmetry with alternating positive and negative signs as we go around the central point $K$ or $K'$. In between we may see a white line of pixels indicating a node. This indicates a $cos(3\phi)$ type of behavior for the envelope function. There appears to be a slight offset of the nodal lines from the symmetry directions but this may arise from an arbitrary overall phase of the two particle eigenstates.   It indicates that if the matrix elements of the velocity operator in the plane are fully symmetric around $K$ as we expect, then the exciton will be dark by this modulation of the three fold symmetry envelope function.  Furthermore we can see that for the $\lambda=3$ exciton the signs of the {\bf k}-space envelope function are opposite near $K$ and $K'$  or more precisely the function is odd with respect to a vertical mirror plane going through the $\Gamma-M$ line while for $\lambda=4$ the signs are the same. So, we conclude that exciton 3 can be described as an $A_1'$ exciton while $\lambda=3$ is an $A_2'$ exciton.  But both become dark by the three-fold symmetry envelope function. 

As already discussed in Sec. \ref{sec:overview} a $\cos{(m\phi)}$ behavior 
is expected but the question arises why we only see a $m=3$ or $f$-like envelope function for these lowest energy dark excitons. 
States with lower $m$ are indeed expected but would be more spread out in real space (the higher the anguar momentum the more localized by the centrifugal term (of the form $m^2/r^2$) and hence more localized in momentum space; therefore, in order to access $p$-like or $d$-like states originating from and around point $K$, as reported in \cite{Qiu16} for MoS$_2$, we would need a much finer {\bf k} mesh to capture the fine structure around this point. With the {\bf k} mesh used in this study, we only access points in momentum space which are not close "enough" to $K$ responsible for $p$- or $d$-like states.

Our analysis in terms of the envelope function resembles the approach used by Qiu \etal \cite{Qiu16} for MoS$_2$. However, they used a special dual mesh approach, which allowed them to use {\bf k} meshes as large as $300\times300$. With our 10 time smaller divisions we still obtain a rather pixelated view of the {\bf k} space behavior. Therefore we can unfortunately not  obtain the nodal structure near each point $K$ as accurate. The effective k-spacing is $(2\pi/a)/30$ is about 0.06 \AA$^{-1}$ but the size of the exciton spread itself is of order 1 \AA$^{-1}$ and thus we don't have sufficient resolution to fully see the fine structure which would reveal the Rydberg like excited states of the exciton. The number of excitons we can resolve also depends strongly on the fineness of the \textbf{k} mesh and as mentioned before 
does not allow us to capture the $p$ or $d$-like excitons. 

In fact, we did notice that the number or excitons  depends on the size of the {\bf k}-mesh and went up to $35\times35\times3$ with $N_v=1$, $N_c=1$ but were still not able to obtain the $p$- or $d$-like 
excitons.

\begin{figure*}
\includegraphics[width=1\textwidth,trim=4.8cm .8cm 4cm 1.1cm, clip]{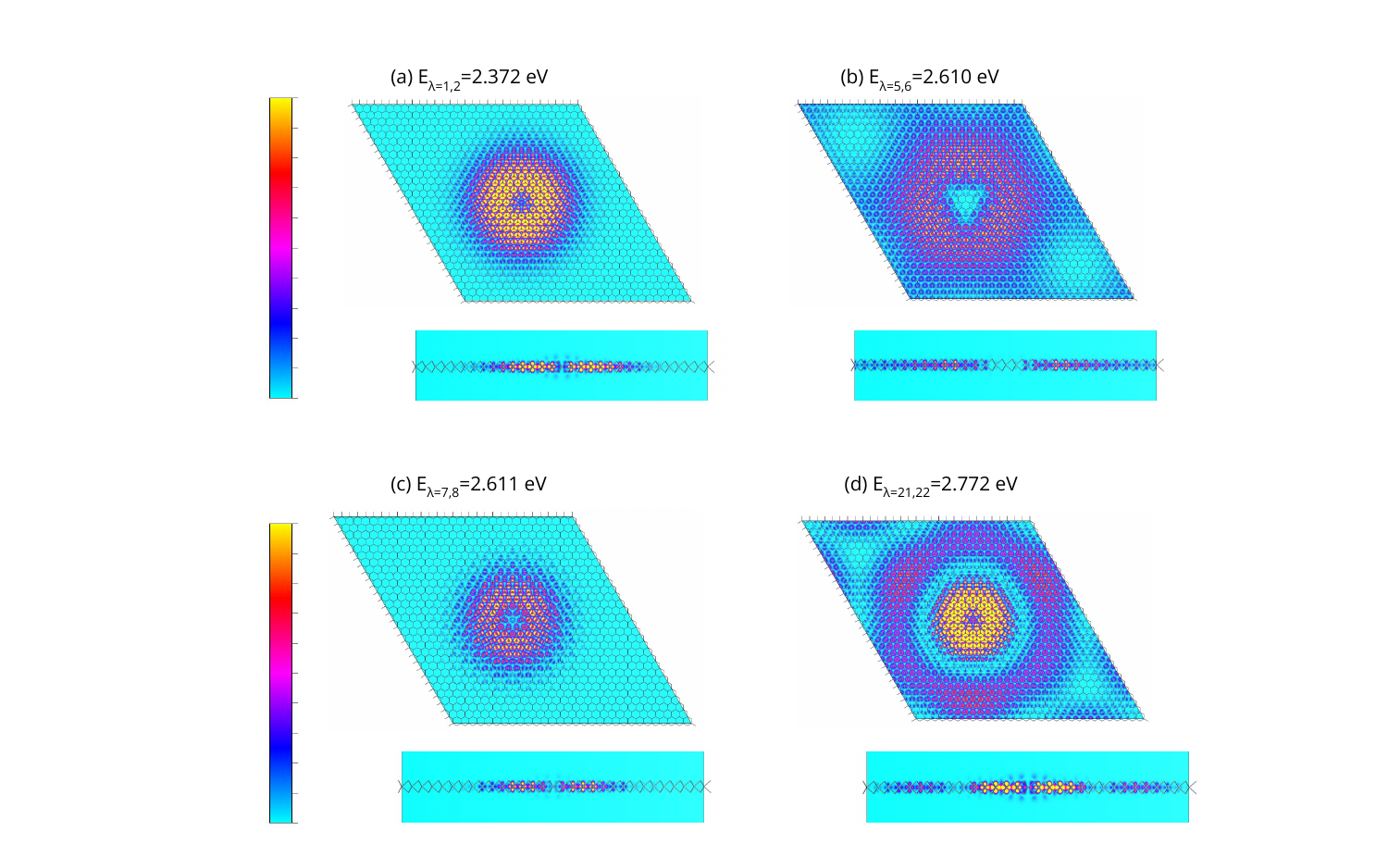}
  \caption{Real-space probability density of the first three bright excitons in energy and one higher energy exciton, shown for a section that goes through the tungsten atoms perpendicular (Top) and parallel (bottom) to the out-of-plane axis, respectively. In the bottom row of each panel, the section is situated in the middle of the suppercell. For the color bar limits, the same minimum (cyan) and maximum (yellow) values are used in all three maps. Values smaller (larger) than this range are shown by the saturation colors cyan (yellow).}
    \label{rbright}
\end{figure*}
\begin{figure*}
\includegraphics[width=1\textwidth,trim=4.8cm 8.5cm 4cm 1.1cm, clip]{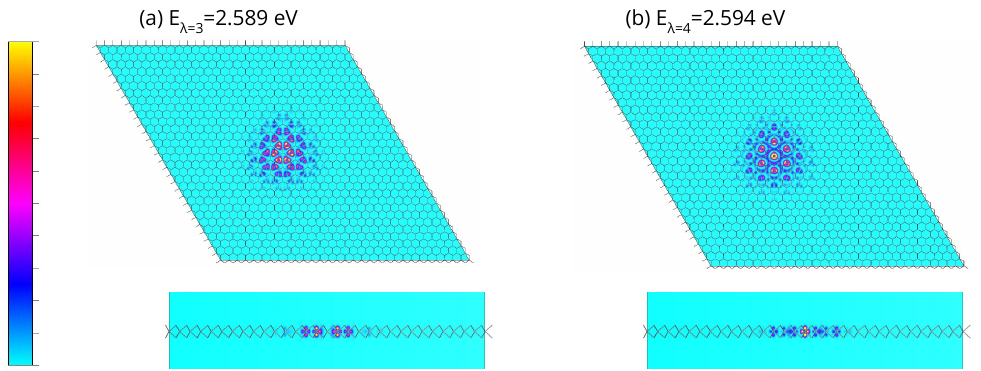}
  \caption{Real-space probability density of the first two dark excitons in energy, shown for a section that goes through the tungsten atoms perpendicular (Top) and parallel (bottom) to the out-of-plane axis, respectively. In the bottom row of each panel, the section is situated in the middle of the supercell. 
  For the color bar limits, the same minimum (cyan) and maximum (yellow) values are used in both maps. Values 
  of $|\Psi|^2$ smaller (larger) than this range are shown by the saturation color cyan (yellow).}
    \label{rdark}
\end{figure*}

\subsubsection{Real-space analysis} \label{sec:realspace}
Next we turn our attention to the real space spread of the exciton wave functions. In Figures \ref{rbright} and \ref{rdark} we visualize the exciton wave functions  in real space as a colormap. The interlayer distance of 25 \AA~is used for this calculation.

Fig. \ref{rbright}(a) shows the spatial distribution of the lowest energy exciton on a section going through the W atom, perpendicular to the c-axis (top) and perpendicular to the a-axis (bottom). As we can see, the overall shape resembles the hexagonal symmetry of the structure rotated by 60$^\circ$  such that the highest weight (yellow) lies on a hexagonal region extending from two to four unit cells away from the hole. Looking closer, we can see that there is a 3-fold symmetry. Also, around each W atom, a dumbbell-like distribution of the electron density can be seen, which originates from the $d$-orbitals of the individual W atoms. This can best be seen in the edge-on projection, which clearly shows $d$-orbital shapes. The orientation of this dumbbell-like distribution changes locally going from one edge to the other. For the two dark excitons we see a different character shown in Fig. \ref{rdark}. In \ref{rdark}(a), the distribution is the highest around W while in \ref{rdark}(b), it is almost zero around W. Furthermore, in Fig. \ref{rdark}(a) we see clear nodal lines at 120$^\circ$ from each other, reminiscent of the $\cos{(3\phi)}$ pattern in the {\bf k}-space figure. 
In Fig. \ref{rdark}(a) we still see regions of low values in rings around each atom but still showing some three-fold symmetry. However, their overall sizes in the real space are pretty close to each other and somewhat smaller than the previous bright exciton. On the other hand, for the next bright exciton shown in Fig. \ref{rbright}(b) is spread out substantially farther. As the exciton energy approaches the gap and the exciton binding energy is reduced, the exciton spreads further in real space. 

Interestingly, even for the lowest energy bright exciton the envelope function is not at its maximum in the center but is rather ring shaped with a dip in the very center. In the second bright exciton, shown in Fig. \ref{rbright}(b),
the central region is distinctly triangular shaped corresponds to a larger central region with very low value. The spatial extent of this exciton is significantly larger than the first bright exciton, consistent with the $2s$ character.
These features can be explained in terms of the 2D envelope functions which correspond to the electron-hole attraction potential. Because of the 2D screening being very distant dependent, it behaves like a weak logarithmic potential $\ln{(r)}$ near the origin and as a unscreened $1/r$ potential far away \cite{Cudazzo11}.  The solutions for the 2D Schr\"odinger equation with  a logarithmic potential were studied by Atabek, Deutsch and Lavaud \cite{Atabek74}. Because of the weak logarithmic potential the $m=0$ (or $s$-like states)  do not start as $r^0$ or a constant but more like $r$ as can be seen in Fig. 4 of \cite{Atabek74}. They go through a single maximum at finite $r$ consistent with the ring shaped envelope we see.   For the $2s$ state, there is a node but the onset starts not at $r=0$ but at finite $r$.  The stronger unscreened Coulomb interaction at large distance means that excited states like $2s$ states are  pushed out to larger $r$. This is consistent with the large triangular hole in this exciton wave function. The triangular shape at close distance in real space results from the triangular warping of the  bands involved in {\bf k} space at large {\bf k}. Even for the lowest exciton, the center region is triangular rather than circular. The next bright exciton, shown in Fig. \ref{rbright}(c) shows very similar behavior as the first bright exciton.
We recall that the doubly degenerate exciton labeled 5,6 is the X$^\prime$ exciton while 7,8 corresponds to  the X$^\dprime$ exciton. (We here call them X instead of A,B because our calculation does not include spin-orbit coupling).
It is somewhat surprising that X$^\dprime$ and X are so similar in spatial extent. However, we can see from Fig. \ref{rbright}(a,c) that they have different probability density. We picked the minimum and maximum values mapped to the color scale the same for both to facilitate a an absolute comparison. They also have somewhat different {\bf k}-space localization: X$^\dprime$ falls of faster  with $k$ away from the band edge $K$ than X. Finally, the bright exciton with largest exciton energy (Fig. \ref{rbright}(d) is seen to have a wider spread and a clear radial node structure. 

Overall, the real space and reciprocal space exciton wave functions show an interesting variety of behaviors which can largely be understood in terms of the 2D problem with a Keldysh type distant dependent screened Coulomb potential.

\section{Conclusions}\label{sec:conclusion}
In this paper we presented a study of the band structure and excitons in monolayer WSe$_2$ using an all-electron muffin-tin-orbital implementation of the QS$GW$ and BSE approaches. The calculations went beyond the usual IPA and RPA based $GW$ approach by including the electron-hole interaction effects in the screening of $W$, which is then called $\hat{W}$. The quasiparticle gaps were found to converge slowly as function of interlayer distance, as $1/d$ and the replacement of $W$ by $\hat{W}$ essentially provides a constant downward shift of the gap. The exciton spectrum was found in excellent agreement with prior theory work \cite{Marsili2021PRB} and various experiments when SOC corrections of the bands are added {\sl a-posteriori}. As expected, the excitons show a smaller variation as function of interlayer distance, as a result of the compensation of the exciton binding energy and quasiparticle gap self-energy shift which are both proportional to $\hat{W}$. Converged exciton eigenvalues were shown to require a very fine in-plane ${\bf k}$ mesh in order to obtain agreement with experimental exciton gaps.  The nature of dark and bright excitons was analyzed in terms of their reciprocal space and real space envelope functions and showed distinct behaviors. First, all the excitons examined were derived almost exclusively from the top valence and lowest conduction bands near $K$. But the dark ones had zero value at $K$ itself. The darkness of two  of the excitons  was found to result from the three-fold symmetry nodal pattern of the envelope function but these two dark excitons still show distinct behavior in real space. The exciton symmetries were analyzed in the context of the Wannier-Mott theory adjusted for 2D. The three-fold symmetry of the first dark excitons found here indicates $f$-like excitons. We expect that there are also $p$- and $d$-like envelope functions in a Rydberg like series, but the reason why the lower  angular momenta $p$- or $d$-like excitons could not be revealed is related to the limitations in the number of {\bf k} points we could include in the BSE. A finer mesh near the band extrema at $K$ and $-K$ is required to capture these excitons which are expected to be more localized in {\bf k} space. Among the bright excitons, we were able to identify the A$'$ exciton and identify it with a $2s$ like exciton state. Our present study provides a detailed look at how the Mott-Wannier model explains qualitatively the exciton series. However, it  shows that the strongly distance dependent screening of the Coulomb interaction between electron and hole modifies  the spatial extent and shape of the exciton envelope functions. These features emerge naturally from the first-principles BSE approach even though we do not explicitly solve for a Wannier-Mott slowly varying envelope function. 

\acknowledgements{This work was supported by the U.S. Department of Energy Basic Energy Sciences (DOE-BES) under Grant No. DE-SC0008933.  Calculations made use of the High Performance Computing Resource in the Core Facility for Advanced Research Computing at Case Western Reserve University.}
\bibliography{lmto,dft,gw,wse2,bse}

\begin{thebibliography}{52}%
\makeatletter
\providecommand \@ifxundefined [1]{%
 \@ifx{#1\undefined}
}%
\providecommand \@ifnum [1]{%
 \ifnum #1\expandafter \@firstoftwo
 \else \expandafter \@secondoftwo
 \fi
}%
\providecommand \@ifx [1]{%
 \ifx #1\expandafter \@firstoftwo
 \else \expandafter \@secondoftwo
 \fi
}%
\providecommand \natexlab [1]{#1}%
\providecommand \enquote  [1]{``#1''}%
\providecommand \bibnamefont  [1]{#1}%
\providecommand \bibfnamefont [1]{#1}%
\providecommand \citenamefont [1]{#1}%
\providecommand \href@noop [0]{\@secondoftwo}%
\providecommand \href [0]{\begingroup \@sanitize@url \@href}%
\providecommand \@href[1]{\@@startlink{#1}\@@href}%
\providecommand \@@href[1]{\endgroup#1\@@endlink}%
\providecommand \@sanitize@url [0]{\catcode `\\12\catcode `\$12\catcode `\&12\catcode `\#12\catcode `\^12\catcode `\_12\catcode `\%12\relax}%
\providecommand \@@startlink[1]{}%
\providecommand \@@endlink[0]{}%
\providecommand \url  [0]{\begingroup\@sanitize@url \@url }%
\providecommand \@url [1]{\endgroup\@href {#1}{\urlprefix }}%
\providecommand \urlprefix  [0]{URL }%
\providecommand \Eprint [0]{\href }%
\providecommand \doibase [0]{https://doi.org/}%
\providecommand \selectlanguage [0]{\@gobble}%
\providecommand \bibinfo  [0]{\@secondoftwo}%
\providecommand \bibfield  [0]{\@secondoftwo}%
\providecommand \translation [1]{[#1]}%
\providecommand \BibitemOpen [0]{}%
\providecommand \bibitemStop [0]{}%
\providecommand \bibitemNoStop [0]{.\EOS\space}%
\providecommand \EOS [0]{\spacefactor3000\relax}%
\providecommand \BibitemShut  [1]{\csname bibitem#1\endcsname}%
\let\auto@bib@innerbib\@empty
\bibitem [{\citenamefont {Keldysh}(1979)}]{Keldysh79}%
  \BibitemOpen
  \bibfield  {author} {\bibinfo {author} {\bibfnamefont {L.~V.}\ \bibnamefont {Keldysh}},\ }\bibfield  {title} {\bibinfo {title} {{Coulomb interaction in thin semiconductor and semimetal films}},\ }\href {http://jetpletters.ru/ps/1458/article_22207.shtml} {\bibfield  {journal} {\bibinfo  {journal} {JETP Lett.}\ }\textbf {\bibinfo {volume} {29}},\ \bibinfo {pages} {658} (\bibinfo {year} {1979})}\BibitemShut {NoStop}%
\bibitem [{\citenamefont {Cudazzo}\ \emph {et~al.}(2011)\citenamefont {Cudazzo}, \citenamefont {Tokatly},\ and\ \citenamefont {Rubio}}]{Cudazzo11}%
  \BibitemOpen
  \bibfield  {author} {\bibinfo {author} {\bibfnamefont {P.}~\bibnamefont {Cudazzo}}, \bibinfo {author} {\bibfnamefont {I.~V.}\ \bibnamefont {Tokatly}},\ and\ \bibinfo {author} {\bibfnamefont {A.}~\bibnamefont {Rubio}},\ }\bibfield  {title} {\bibinfo {title} {{Dielectric screening in two-dimensional insulators: Implications for excitonic and impurity states in graphane}},\ }\href {https://doi.org/10.1103/PhysRevB.84.085406} {\bibfield  {journal} {\bibinfo  {journal} {Phys. Rev. B}\ }\textbf {\bibinfo {volume} {84}},\ \bibinfo {pages} {085406} (\bibinfo {year} {2011})}\BibitemShut {NoStop}%
\bibitem [{\citenamefont {Shinada}\ and\ \citenamefont {Sugano}(1966)}]{Shinada66}%
  \BibitemOpen
  \bibfield  {author} {\bibinfo {author} {\bibfnamefont {M.}~\bibnamefont {Shinada}}\ and\ \bibinfo {author} {\bibfnamefont {S.}~\bibnamefont {Sugano}},\ }\bibfield  {title} {\bibinfo {title} {{Interband Optical Transitions in Extremely Anisotropic Semiconductors. I. Bound and Unbound Exciton Absorption}},\ }\href {https://doi.org/10.1143/JPSJ.21.1936} {\bibfield  {journal} {\bibinfo  {journal} {Journal of the Physical Society of Japan}\ }\textbf {\bibinfo {volume} {21}},\ \bibinfo {pages} {1936} (\bibinfo {year} {1966})}\BibitemShut {NoStop}%
\bibitem [{\citenamefont {Delerue}\ \emph {et~al.}(2003)\citenamefont {Delerue}, \citenamefont {Allan},\ and\ \citenamefont {Lannoo}}]{Delerue03}%
  \BibitemOpen
  \bibfield  {author} {\bibinfo {author} {\bibfnamefont {C.}~\bibnamefont {Delerue}}, \bibinfo {author} {\bibfnamefont {G.}~\bibnamefont {Allan}},\ and\ \bibinfo {author} {\bibfnamefont {M.}~\bibnamefont {Lannoo}},\ }\bibfield  {title} {\bibinfo {title} {{Dimensionality-Dependent Self-Energy Corrections and Exchange-Correlation Potential in Semiconductor Nanostructures}},\ }\href {https://doi.org/10.1103/PhysRevLett.90.076803} {\bibfield  {journal} {\bibinfo  {journal} {Phys. Rev. Lett.}\ }\textbf {\bibinfo {volume} {90}},\ \bibinfo {pages} {076803} (\bibinfo {year} {2003})}\BibitemShut {NoStop}%
\bibitem [{\citenamefont {Komsa}\ and\ \citenamefont {Krasheninnikov}(2012)}]{Komsa12}%
  \BibitemOpen
  \bibfield  {author} {\bibinfo {author} {\bibfnamefont {H.-P.}\ \bibnamefont {Komsa}}\ and\ \bibinfo {author} {\bibfnamefont {A.~V.}\ \bibnamefont {Krasheninnikov}},\ }\bibfield  {title} {\bibinfo {title} {{Effects of confinement and environment on the electronic structure and exciton binding energy of MoS${}_{2}$ from first principles}},\ }\href {https://doi.org/10.1103/PhysRevB.86.241201} {\bibfield  {journal} {\bibinfo  {journal} {Phys. Rev. B}\ }\textbf {\bibinfo {volume} {86}},\ \bibinfo {pages} {241201} (\bibinfo {year} {2012})}\BibitemShut {NoStop}%
\bibitem [{\citenamefont {Hedin}(1965)}]{Hedin65}%
  \BibitemOpen
  \bibfield  {author} {\bibinfo {author} {\bibfnamefont {L.}~\bibnamefont {Hedin}},\ }\bibfield  {title} {\bibinfo {title} {New method for calculating the one-particle green's function with application to the electron-gas problem},\ }\href {https://doi.org/10.1103/PhysRev.139.A796} {\bibfield  {journal} {\bibinfo  {journal} {Phys. Rev.}\ }\textbf {\bibinfo {volume} {139}},\ \bibinfo {pages} {A796} (\bibinfo {year} {1965})}\BibitemShut {NoStop}%
\bibitem [{\citenamefont {Hedin}\ and\ \citenamefont {Lundqvist}(1969)}]{Hedin69}%
  \BibitemOpen
  \bibfield  {author} {\bibinfo {author} {\bibfnamefont {L.}~\bibnamefont {Hedin}}\ and\ \bibinfo {author} {\bibfnamefont {S.}~\bibnamefont {Lundqvist}},\ }\bibfield  {title} {\bibinfo {title} {Effects of electron-electron and electron-phonon interactions on the one-electron states of solids},\ }in\ \href@noop {} {\emph {\bibinfo {booktitle} {Solid State Physics, Advanced in Research and Applications}}},\ Vol.~\bibinfo {volume} {23},\ \bibinfo {editor} {edited by\ \bibinfo {editor} {\bibfnamefont {F.}~\bibnamefont {Seitz}}, \bibinfo {editor} {\bibfnamefont {D.}~\bibnamefont {Turnbull}},\ and\ \bibinfo {editor} {\bibfnamefont {H.}~\bibnamefont {Ehrenreich}}}\ (\bibinfo  {publisher} {Academic Press},\ \bibinfo {address} {New York},\ \bibinfo {year} {1969})\ pp.\ \bibinfo {pages} {1--181}\BibitemShut {NoStop}%
\bibitem [{\citenamefont {Strinati}(1988)}]{Strinati88}%
  \BibitemOpen
  \bibfield  {author} {\bibinfo {author} {\bibfnamefont {G.}~\bibnamefont {Strinati}},\ }\bibfield  {title} {\bibinfo {title} {Application of the green's functions method to the study of the optical properties of semiconductors},\ }\href {https://doi.org/10.1007/BF02725962} {\bibfield  {journal} {\bibinfo  {journal} {La Rivista del Nuovo Cimento (1978-1999)}\ }\textbf {\bibinfo {volume} {11}},\ \bibinfo {pages} {1} (\bibinfo {year} {1988})}\BibitemShut {NoStop}%
\bibitem [{\citenamefont {Hanke}(1978)}]{Hanke78}%
  \BibitemOpen
  \bibfield  {author} {\bibinfo {author} {\bibfnamefont {W.}~\bibnamefont {Hanke}},\ }\bibfield  {title} {\bibinfo {title} {{Dielectric theory of elementary excitations in crystals}},\ }\href {https://doi.org/10.1080/00018737800101384} {\bibfield  {journal} {\bibinfo  {journal} {Advances in Physics}\ }\textbf {\bibinfo {volume} {27}},\ \bibinfo {pages} {287} (\bibinfo {year} {1978})}\BibitemShut {NoStop}%
\bibitem [{\citenamefont {Hanke}\ and\ \citenamefont {Sham}(1980)}]{HankeSham80}%
  \BibitemOpen
  \bibfield  {author} {\bibinfo {author} {\bibfnamefont {W.}~\bibnamefont {Hanke}}\ and\ \bibinfo {author} {\bibfnamefont {L.~J.}\ \bibnamefont {Sham}},\ }\bibfield  {title} {\bibinfo {title} {{Many-particle effects in the optical spectrum of a semiconductor}},\ }\href {https://doi.org/10.1103/PhysRevB.21.4656} {\bibfield  {journal} {\bibinfo  {journal} {Phys. Rev. B}\ }\textbf {\bibinfo {volume} {21}},\ \bibinfo {pages} {4656} (\bibinfo {year} {1980})}\BibitemShut {NoStop}%
\bibitem [{\citenamefont {Albrecht}\ \emph {et~al.}(1998)\citenamefont {Albrecht}, \citenamefont {Reining}, \citenamefont {Del~Sole},\ and\ \citenamefont {Onida}}]{Albrecht98}%
  \BibitemOpen
  \bibfield  {author} {\bibinfo {author} {\bibfnamefont {S.}~\bibnamefont {Albrecht}}, \bibinfo {author} {\bibfnamefont {L.}~\bibnamefont {Reining}}, \bibinfo {author} {\bibfnamefont {R.}~\bibnamefont {Del~Sole}},\ and\ \bibinfo {author} {\bibfnamefont {G.}~\bibnamefont {Onida}},\ }\bibfield  {title} {\bibinfo {title} {Ab initio calculation of excitonic effects in the optical spectra of semiconductors},\ }\href {https://doi.org/10.1103/PhysRevLett.80.4510} {\bibfield  {journal} {\bibinfo  {journal} {Phys. Rev. Lett.}\ }\textbf {\bibinfo {volume} {80}},\ \bibinfo {pages} {4510} (\bibinfo {year} {1998})}\BibitemShut {NoStop}%
\bibitem [{\citenamefont {Rohlfing}\ and\ \citenamefont {Louie}(1998)}]{Rohlfing98}%
  \BibitemOpen
  \bibfield  {author} {\bibinfo {author} {\bibfnamefont {M.}~\bibnamefont {Rohlfing}}\ and\ \bibinfo {author} {\bibfnamefont {S.~G.}\ \bibnamefont {Louie}},\ }\bibfield  {title} {\bibinfo {title} {{Electron-Hole Excitations in Semiconductors and Insulators}},\ }\href {https://doi.org/10.1103/PhysRevLett.81.2312} {\bibfield  {journal} {\bibinfo  {journal} {Phys. Rev. Lett.}\ }\textbf {\bibinfo {volume} {81}},\ \bibinfo {pages} {2312} (\bibinfo {year} {1998})}\BibitemShut {NoStop}%
\bibitem [{\citenamefont {Benedict}\ \emph {et~al.}(1998)\citenamefont {Benedict}, \citenamefont {Shirley},\ and\ \citenamefont {Bohn}}]{Benedict98}%
  \BibitemOpen
  \bibfield  {author} {\bibinfo {author} {\bibfnamefont {L.~X.}\ \bibnamefont {Benedict}}, \bibinfo {author} {\bibfnamefont {E.~L.}\ \bibnamefont {Shirley}},\ and\ \bibinfo {author} {\bibfnamefont {R.~B.}\ \bibnamefont {Bohn}},\ }\bibfield  {title} {\bibinfo {title} {Optical absorption of insulators and the electron-hole interaction: An ab initio calculation},\ }\href {https://doi.org/10.1103/PhysRevLett.80.4514} {\bibfield  {journal} {\bibinfo  {journal} {Phys. Rev. Lett.}\ }\textbf {\bibinfo {volume} {80}},\ \bibinfo {pages} {4514} (\bibinfo {year} {1998})}\BibitemShut {NoStop}%
\bibitem [{\citenamefont {Onida}\ \emph {et~al.}(2002)\citenamefont {Onida}, \citenamefont {Reining},\ and\ \citenamefont {Rubio}}]{Onida02}%
  \BibitemOpen
  \bibfield  {author} {\bibinfo {author} {\bibfnamefont {G.}~\bibnamefont {Onida}}, \bibinfo {author} {\bibfnamefont {L.}~\bibnamefont {Reining}},\ and\ \bibinfo {author} {\bibfnamefont {A.}~\bibnamefont {Rubio}},\ }\bibfield  {title} {\bibinfo {title} {{Electronic excitations: density-functional versus many-body Green's-function approaches}},\ }\href {https://doi.org/10.1103/RevModPhys.74.601} {\bibfield  {journal} {\bibinfo  {journal} {Rev. Mod. Phys.}\ }\textbf {\bibinfo {volume} {74}},\ \bibinfo {pages} {601} (\bibinfo {year} {2002})}\BibitemShut {NoStop}%
\bibitem [{\citenamefont {Kotani}\ \emph {et~al.}(2007)\citenamefont {Kotani}, \citenamefont {van Schilfgaarde},\ and\ \citenamefont {Faleev}}]{Kotani07}%
  \BibitemOpen
  \bibfield  {author} {\bibinfo {author} {\bibfnamefont {T.}~\bibnamefont {Kotani}}, \bibinfo {author} {\bibfnamefont {M.}~\bibnamefont {van Schilfgaarde}},\ and\ \bibinfo {author} {\bibfnamefont {S.~V.}\ \bibnamefont {Faleev}},\ }\bibfield  {title} {\bibinfo {title} {{Quasiparticle self-consistent {GW} method: A basis for the independent-particle approximation}},\ }\href {https://doi.org/10.1103/PhysRevB.76.165106} {\bibfield  {journal} {\bibinfo  {journal} {Phys.Rev. B}\ }\textbf {\bibinfo {volume} {76}},\ \bibinfo {eid} {165106} (\bibinfo {year} {2007})}\BibitemShut {NoStop}%
\bibitem [{\citenamefont {Pashov}\ \emph {et~al.}(2019)\citenamefont {Pashov}, \citenamefont {Acharya}, \citenamefont {Lambrecht}, \citenamefont {Jackson}, \citenamefont {Belashchenko}, \citenamefont {Chantis}, \citenamefont {Jamet},\ and\ \citenamefont {van Schilfgaarde}}]{questaalpaper}%
  \BibitemOpen
  \bibfield  {author} {\bibinfo {author} {\bibfnamefont {D.}~\bibnamefont {Pashov}}, \bibinfo {author} {\bibfnamefont {S.}~\bibnamefont {Acharya}}, \bibinfo {author} {\bibfnamefont {W.~R.}\ \bibnamefont {Lambrecht}}, \bibinfo {author} {\bibfnamefont {J.}~\bibnamefont {Jackson}}, \bibinfo {author} {\bibfnamefont {K.~D.}\ \bibnamefont {Belashchenko}}, \bibinfo {author} {\bibfnamefont {A.}~\bibnamefont {Chantis}}, \bibinfo {author} {\bibfnamefont {F.}~\bibnamefont {Jamet}},\ and\ \bibinfo {author} {\bibfnamefont {M.}~\bibnamefont {van Schilfgaarde}},\ }\bibfield  {title} {\bibinfo {title} {{Questaal: A package of electronic structure methods based on the linear muffin-tin orbital technique}},\ }\href {https://doi.org/https://doi.org/10.1016/j.cpc.2019.107065} {\bibfield  {journal} {\bibinfo  {journal} {Computer Physics Communications}\ ,\ \bibinfo {pages} {107065}} (\bibinfo {year} {2019})}\BibitemShut {NoStop}%
\bibitem [{\citenamefont {Aryasetiawan}\ and\ \citenamefont {Gunnarsson}(1994)}]{Aryasetiawan94}%
  \BibitemOpen
  \bibfield  {author} {\bibinfo {author} {\bibfnamefont {F.}~\bibnamefont {Aryasetiawan}}\ and\ \bibinfo {author} {\bibfnamefont {O.}~\bibnamefont {Gunnarsson}},\ }\bibfield  {title} {\bibinfo {title} {Product-basis method for calculating dielectric matrices},\ }\href {https://doi.org/10.1103/PhysRevB.49.16214} {\bibfield  {journal} {\bibinfo  {journal} {Phys. Rev. B}\ }\textbf {\bibinfo {volume} {49}},\ \bibinfo {pages} {16214} (\bibinfo {year} {1994})}\BibitemShut {NoStop}%
\bibitem [{\citenamefont {van Schilfgaarde}\ \emph {et~al.}(2006)\citenamefont {van Schilfgaarde}, \citenamefont {Kotani},\ and\ \citenamefont {Faleev}}]{MvSQSGWprl}%
  \BibitemOpen
  \bibfield  {author} {\bibinfo {author} {\bibfnamefont {M.}~\bibnamefont {van Schilfgaarde}}, \bibinfo {author} {\bibfnamefont {T.}~\bibnamefont {Kotani}},\ and\ \bibinfo {author} {\bibfnamefont {S.}~\bibnamefont {Faleev}},\ }\bibfield  {title} {\bibinfo {title} {{Quasiparticle Self-Consistent $GW$ Theory}},\ }\href {https://doi.org/10.1103/PhysRevLett.96.226402} {\bibfield  {journal} {\bibinfo  {journal} {Phys. Rev. Lett.}\ }\textbf {\bibinfo {volume} {96}},\ \bibinfo {pages} {226402} (\bibinfo {year} {2006})}\BibitemShut {NoStop}%
\bibitem [{\citenamefont {Bhandari}\ \emph {et~al.}(2018)\citenamefont {Bhandari}, \citenamefont {van Schilfgaarde}, \citenamefont {Kotani},\ and\ \citenamefont {Lambrecht}}]{Bhandari18}%
  \BibitemOpen
  \bibfield  {author} {\bibinfo {author} {\bibfnamefont {C.}~\bibnamefont {Bhandari}}, \bibinfo {author} {\bibfnamefont {M.}~\bibnamefont {van Schilfgaarde}}, \bibinfo {author} {\bibfnamefont {T.}~\bibnamefont {Kotani}},\ and\ \bibinfo {author} {\bibfnamefont {W.~R.~L.}\ \bibnamefont {Lambrecht}},\ }\bibfield  {title} {\bibinfo {title} {{All-electron quasiparticle self-consistent $\mathit{GW}$ band structures for ${\mathrm{SrTiO}}_{3}$ including lattice polarization corrections in different phases}},\ }\href {https://doi.org/10.1103/PhysRevMaterials.2.013807} {\bibfield  {journal} {\bibinfo  {journal} {Phys. Rev. Materials}\ }\textbf {\bibinfo {volume} {2}},\ \bibinfo {pages} {013807} (\bibinfo {year} {2018})}\BibitemShut {NoStop}%
\bibitem [{\citenamefont {Shishkin}\ \emph {et~al.}(2007)\citenamefont {Shishkin}, \citenamefont {Marsman},\ and\ \citenamefont {Kresse}}]{Shishkin07}%
  \BibitemOpen
  \bibfield  {author} {\bibinfo {author} {\bibfnamefont {M.}~\bibnamefont {Shishkin}}, \bibinfo {author} {\bibfnamefont {M.}~\bibnamefont {Marsman}},\ and\ \bibinfo {author} {\bibfnamefont {G.}~\bibnamefont {Kresse}},\ }\bibfield  {title} {\bibinfo {title} {{Accurate Quasiparticle Spectra from Self-Consistent GW Calculations with Vertex Corrections}},\ }\href {https://doi.org/10.1103/PhysRevLett.99.246403} {\bibfield  {journal} {\bibinfo  {journal} {Phys. Rev. Lett.}\ }\textbf {\bibinfo {volume} {99}},\ \bibinfo {pages} {246403} (\bibinfo {year} {2007})}\BibitemShut {NoStop}%
\bibitem [{\citenamefont {Chen}\ and\ \citenamefont {Pasquarello}(2015)}]{ChenPasquarello15}%
  \BibitemOpen
  \bibfield  {author} {\bibinfo {author} {\bibfnamefont {W.}~\bibnamefont {Chen}}\ and\ \bibinfo {author} {\bibfnamefont {A.}~\bibnamefont {Pasquarello}},\ }\bibfield  {title} {\bibinfo {title} {{Accurate band gaps of extended systems via efficient vertex corrections in $GW$}},\ }\href {https://doi.org/10.1103/PhysRevB.92.041115} {\bibfield  {journal} {\bibinfo  {journal} {Phys. Rev. B}\ }\textbf {\bibinfo {volume} {92}},\ \bibinfo {pages} {041115} (\bibinfo {year} {2015})}\BibitemShut {NoStop}%
\bibitem [{\citenamefont {Cunningham}\ \emph {et~al.}(2018)\citenamefont {Cunningham}, \citenamefont {Gr\"uning}, \citenamefont {Azarhoosh}, \citenamefont {Pashov},\ and\ \citenamefont {van Schilfgaarde}}]{Cunningham18}%
  \BibitemOpen
  \bibfield  {author} {\bibinfo {author} {\bibfnamefont {B.}~\bibnamefont {Cunningham}}, \bibinfo {author} {\bibfnamefont {M.}~\bibnamefont {Gr\"uning}}, \bibinfo {author} {\bibfnamefont {P.}~\bibnamefont {Azarhoosh}}, \bibinfo {author} {\bibfnamefont {D.}~\bibnamefont {Pashov}},\ and\ \bibinfo {author} {\bibfnamefont {M.}~\bibnamefont {van Schilfgaarde}},\ }\bibfield  {title} {\bibinfo {title} {{Effect of ladder diagrams on optical absorption spectra in a quasiparticle self-consistent $\mathit{GW}$ framework}},\ }\href {https://doi.org/10.1103/PhysRevMaterials.2.034603} {\bibfield  {journal} {\bibinfo  {journal} {Phys. Rev. Materials}\ }\textbf {\bibinfo {volume} {2}},\ \bibinfo {pages} {034603} (\bibinfo {year} {2018})}\BibitemShut {NoStop}%
\bibitem [{\citenamefont {Cunningham}\ \emph {et~al.}(2023)\citenamefont {Cunningham}, \citenamefont {Gr\"uning}, \citenamefont {Pashov},\ and\ \citenamefont {van Schilfgaarde}}]{Cunningham23}%
  \BibitemOpen
  \bibfield  {author} {\bibinfo {author} {\bibfnamefont {B.}~\bibnamefont {Cunningham}}, \bibinfo {author} {\bibfnamefont {M.}~\bibnamefont {Gr\"uning}}, \bibinfo {author} {\bibfnamefont {D.}~\bibnamefont {Pashov}},\ and\ \bibinfo {author} {\bibfnamefont {M.}~\bibnamefont {van Schilfgaarde}},\ }\bibfield  {title} {\bibinfo {title} {{$\mathrm{QS}G\hat{W}$: Quasiparticle self-consistent $GW$ with ladder diagrams in $W$}},\ }\href {https://doi.org/10.1103/PhysRevB.108.165104} {\bibfield  {journal} {\bibinfo  {journal} {Phys. Rev. B}\ }\textbf {\bibinfo {volume} {108}},\ \bibinfo {pages} {165104} (\bibinfo {year} {2023})}\BibitemShut {NoStop}%
\bibitem [{\citenamefont {Radha}\ \emph {et~al.}(2021)\citenamefont {Radha}, \citenamefont {Lambrecht}, \citenamefont {Cunningham}, \citenamefont {Gr\"uning}, \citenamefont {Pashov},\ and\ \citenamefont {van Schilfgaarde}}]{Radha21}%
  \BibitemOpen
  \bibfield  {author} {\bibinfo {author} {\bibfnamefont {S.~K.}\ \bibnamefont {Radha}}, \bibinfo {author} {\bibfnamefont {W.~R.~L.}\ \bibnamefont {Lambrecht}}, \bibinfo {author} {\bibfnamefont {B.}~\bibnamefont {Cunningham}}, \bibinfo {author} {\bibfnamefont {M.}~\bibnamefont {Gr\"uning}}, \bibinfo {author} {\bibfnamefont {D.}~\bibnamefont {Pashov}},\ and\ \bibinfo {author} {\bibfnamefont {M.}~\bibnamefont {van Schilfgaarde}},\ }\bibfield  {title} {\bibinfo {title} {{Optical response and band structure of ${\mathrm{LiCoO}}_{2}$ including electron-hole interaction effects}},\ }\href {https://doi.org/10.1103/PhysRevB.104.115120} {\bibfield  {journal} {\bibinfo  {journal} {Phys. Rev. B}\ }\textbf {\bibinfo {volume} {104}},\ \bibinfo {pages} {115120} (\bibinfo {year} {2021})}\BibitemShut {NoStop}%
\bibitem [{\citenamefont {Jain}\ \emph {et~al.}(2013)\citenamefont {Jain}, \citenamefont {Ong}, \citenamefont {Hautier}, \citenamefont {Chen}, \citenamefont {Richards}, \citenamefont {Dacek}, \citenamefont {Cholia}, \citenamefont {Gunter}, \citenamefont {Skinner}, \citenamefont {Ceder},\ and\ \citenamefont {Persson}}]{MaterialsProject}%
  \BibitemOpen
  \bibfield  {author} {\bibinfo {author} {\bibfnamefont {A.}~\bibnamefont {Jain}}, \bibinfo {author} {\bibfnamefont {S.~P.}\ \bibnamefont {Ong}}, \bibinfo {author} {\bibfnamefont {G.}~\bibnamefont {Hautier}}, \bibinfo {author} {\bibfnamefont {W.}~\bibnamefont {Chen}}, \bibinfo {author} {\bibfnamefont {W.~D.}\ \bibnamefont {Richards}}, \bibinfo {author} {\bibfnamefont {S.}~\bibnamefont {Dacek}}, \bibinfo {author} {\bibfnamefont {S.}~\bibnamefont {Cholia}}, \bibinfo {author} {\bibfnamefont {D.}~\bibnamefont {Gunter}}, \bibinfo {author} {\bibfnamefont {D.}~\bibnamefont {Skinner}}, \bibinfo {author} {\bibfnamefont {G.}~\bibnamefont {Ceder}},\ and\ \bibinfo {author} {\bibfnamefont {K.~A.}\ \bibnamefont {Persson}},\ }\bibfield  {title} {\bibinfo {title} {{Commentary: The Materials Project: A materials genome approach to accelerating materials innovation}},\ }\href {https://doi.org/10.1063/1.4812323} {\bibfield  {journal} {\bibinfo  {journal} {APL Materials}\ }\textbf {\bibinfo {volume} {1}},\ \bibinfo {pages}
  {011002} (\bibinfo {year} {2013})}\BibitemShut {NoStop}%
\bibitem [{MPw()}]{MPwse2}%
  \BibitemOpen
  \href {https://next-gen.materialsproject.org/materials/mp-1023936?formula=WSe2} {}\bibinfo {note} {Materials Project, mp-1023936}\BibitemShut {NoStop}%
\bibitem [{\citenamefont {Deguchi}\ \emph {et~al.}(2016)\citenamefont {Deguchi}, \citenamefont {Sato}, \citenamefont {Kino},\ and\ \citenamefont {Kotani}}]{Deguchi_2016}%
  \BibitemOpen
  \bibfield  {author} {\bibinfo {author} {\bibfnamefont {D.}~\bibnamefont {Deguchi}}, \bibinfo {author} {\bibfnamefont {K.}~\bibnamefont {Sato}}, \bibinfo {author} {\bibfnamefont {H.}~\bibnamefont {Kino}},\ and\ \bibinfo {author} {\bibfnamefont {T.}~\bibnamefont {Kotani}},\ }\bibfield  {title} {\bibinfo {title} {Accurate energy bands calculated by the hybrid quasiparticle self-consistent gw method implemented in the ecalj package},\ }\href {https://doi.org/10.7567/JJAP.55.051201} {\bibfield  {journal} {\bibinfo  {journal} {Japanese Journal of Applied Physics}\ }\textbf {\bibinfo {volume} {55}},\ \bibinfo {pages} {051201} (\bibinfo {year} {2016})}\BibitemShut {NoStop}%
\bibitem [{\citenamefont {Sander}\ \emph {et~al.}(2015)\citenamefont {Sander}, \citenamefont {Maggio},\ and\ \citenamefont {Kresse}}]{Sander2015}%
  \BibitemOpen
  \bibfield  {author} {\bibinfo {author} {\bibfnamefont {T.}~\bibnamefont {Sander}}, \bibinfo {author} {\bibfnamefont {E.}~\bibnamefont {Maggio}},\ and\ \bibinfo {author} {\bibfnamefont {G.}~\bibnamefont {Kresse}},\ }\bibfield  {title} {\bibinfo {title} {{Beyond the Tamm-Dancoff approximation for extended systems using exact diagonalization}},\ }\href {https://doi.org/10.1103/PhysRevB.92.045209} {\bibfield  {journal} {\bibinfo  {journal} {Phys. Rev. B}\ }\textbf {\bibinfo {volume} {92}},\ \bibinfo {pages} {045209} (\bibinfo {year} {2015})}\BibitemShut {NoStop}%
\bibitem [{\citenamefont {Cheiwchanchamnangij}\ and\ \citenamefont {Lambrecht}(2012)}]{Tawinan12}%
  \BibitemOpen
  \bibfield  {author} {\bibinfo {author} {\bibfnamefont {T.}~\bibnamefont {Cheiwchanchamnangij}}\ and\ \bibinfo {author} {\bibfnamefont {W.~R.~L.}\ \bibnamefont {Lambrecht}},\ }\bibfield  {title} {\bibinfo {title} {{Quasiparticle band structure calculation of monolayer, bilayer, and bulk MoS${}_{2}$}},\ }\href {https://doi.org/10.1103/PhysRevB.85.205302} {\bibfield  {journal} {\bibinfo  {journal} {Phys. Rev. B}\ }\textbf {\bibinfo {volume} {85}},\ \bibinfo {pages} {205302} (\bibinfo {year} {2012})}\BibitemShut {NoStop}%
\bibitem [{\citenamefont {Marsili}\ \emph {et~al.}(2021)\citenamefont {Marsili}, \citenamefont {Molina-S\'anchez}, \citenamefont {Palummo}, \citenamefont {Sangalli},\ and\ \citenamefont {Marini}}]{Marsili2021PRB}%
  \BibitemOpen
  \bibfield  {author} {\bibinfo {author} {\bibfnamefont {M.}~\bibnamefont {Marsili}}, \bibinfo {author} {\bibfnamefont {A.}~\bibnamefont {Molina-S\'anchez}}, \bibinfo {author} {\bibfnamefont {M.}~\bibnamefont {Palummo}}, \bibinfo {author} {\bibfnamefont {D.}~\bibnamefont {Sangalli}},\ and\ \bibinfo {author} {\bibfnamefont {A.}~\bibnamefont {Marini}},\ }\bibfield  {title} {\bibinfo {title} {Spinorial formulation of the $gw$-bse equations and spin properties of excitons in two-dimensional transition metal dichalcogenides},\ }\href {https://doi.org/10.1103/PhysRevB.103.155152} {\bibfield  {journal} {\bibinfo  {journal} {Phys. Rev. B}\ }\textbf {\bibinfo {volume} {103}},\ \bibinfo {pages} {155152} (\bibinfo {year} {2021})}\BibitemShut {NoStop}%
\bibitem [{\citenamefont {Ramasubramaniam}(2012)}]{Ramasubramaniam12}%
  \BibitemOpen
  \bibfield  {author} {\bibinfo {author} {\bibfnamefont {A.}~\bibnamefont {Ramasubramaniam}},\ }\bibfield  {title} {\bibinfo {title} {{Large excitonic effects in monolayers of molybdenum and tungsten dichalcogenides}},\ }\href {https://doi.org/10.1103/PhysRevB.86.115409} {\bibfield  {journal} {\bibinfo  {journal} {Phys. Rev. B}\ }\textbf {\bibinfo {volume} {86}},\ \bibinfo {pages} {115409} (\bibinfo {year} {2012})}\BibitemShut {NoStop}%
\bibitem [{\citenamefont {Qiu}\ \emph {et~al.}(2013)\citenamefont {Qiu}, \citenamefont {da~Jornada},\ and\ \citenamefont {Louie}}]{Qiu13}%
  \BibitemOpen
  \bibfield  {author} {\bibinfo {author} {\bibfnamefont {D.~Y.}\ \bibnamefont {Qiu}}, \bibinfo {author} {\bibfnamefont {F.~H.}\ \bibnamefont {da~Jornada}},\ and\ \bibinfo {author} {\bibfnamefont {S.~G.}\ \bibnamefont {Louie}},\ }\bibfield  {title} {\bibinfo {title} {{Optical Spectrum of ${\mathrm{MoS}}_{2}$: Many-Body Effects and Diversity of Exciton States}},\ }\href {https://doi.org/10.1103/PhysRevLett.111.216805} {\bibfield  {journal} {\bibinfo  {journal} {Phys. Rev. Lett.}\ }\textbf {\bibinfo {volume} {111}},\ \bibinfo {pages} {216805} (\bibinfo {year} {2013})}\BibitemShut {NoStop}%
\bibitem [{\citenamefont {Qiu}\ \emph {et~al.}(2016)\citenamefont {Qiu}, \citenamefont {da~Jornada},\ and\ \citenamefont {Louie}}]{Qiu16}%
  \BibitemOpen
  \bibfield  {author} {\bibinfo {author} {\bibfnamefont {D.~Y.}\ \bibnamefont {Qiu}}, \bibinfo {author} {\bibfnamefont {F.~H.}\ \bibnamefont {da~Jornada}},\ and\ \bibinfo {author} {\bibfnamefont {S.~G.}\ \bibnamefont {Louie}},\ }\bibfield  {title} {\bibinfo {title} {{Screening and many-body effects in two-dimensional crystals: Monolayer ${\mathrm{MoS}}_{2}$}},\ }\href {https://doi.org/10.1103/PhysRevB.93.235435} {\bibfield  {journal} {\bibinfo  {journal} {Phys. Rev. B}\ }\textbf {\bibinfo {volume} {93}},\ \bibinfo {pages} {235435} (\bibinfo {year} {2016})}\BibitemShut {NoStop}%
\bibitem [{\citenamefont {Chen}\ \emph {et~al.}(2019)\citenamefont {Chen}, \citenamefont {Lu}, \citenamefont {Goldstein}, \citenamefont {Tong}, \citenamefont {Chaves}, \citenamefont {Kunstmann}, \citenamefont {Cavalcante}, \citenamefont {Woźniak}, \citenamefont {Seifert}, \citenamefont {Reichman}, \citenamefont {Taniguchi}, \citenamefont {Watanabe}, \citenamefont {Smirnov},\ and\ \citenamefont {Yan}}]{Chen2019NanoLett}%
  \BibitemOpen
  \bibfield  {author} {\bibinfo {author} {\bibfnamefont {S.-Y.}\ \bibnamefont {Chen}}, \bibinfo {author} {\bibfnamefont {Z.}~\bibnamefont {Lu}}, \bibinfo {author} {\bibfnamefont {T.}~\bibnamefont {Goldstein}}, \bibinfo {author} {\bibfnamefont {J.}~\bibnamefont {Tong}}, \bibinfo {author} {\bibfnamefont {A.}~\bibnamefont {Chaves}}, \bibinfo {author} {\bibfnamefont {J.}~\bibnamefont {Kunstmann}}, \bibinfo {author} {\bibfnamefont {L.~S.~R.}\ \bibnamefont {Cavalcante}}, \bibinfo {author} {\bibfnamefont {T.}~\bibnamefont {Woźniak}}, \bibinfo {author} {\bibfnamefont {G.}~\bibnamefont {Seifert}}, \bibinfo {author} {\bibfnamefont {D.~R.}\ \bibnamefont {Reichman}}, \bibinfo {author} {\bibfnamefont {T.}~\bibnamefont {Taniguchi}}, \bibinfo {author} {\bibfnamefont {K.}~\bibnamefont {Watanabe}}, \bibinfo {author} {\bibfnamefont {D.}~\bibnamefont {Smirnov}},\ and\ \bibinfo {author} {\bibfnamefont {J.}~\bibnamefont {Yan}},\ }\bibfield  {title} {\bibinfo {title} {{Luminescent Emission of Excited Rydberg Excitons from
  Monolayer WSe$_2$}},\ }\href {https://doi.org/10.1021/acs.nanolett.9b00029} {\bibfield  {journal} {\bibinfo  {journal} {Nano Letters}\ }\textbf {\bibinfo {volume} {19}},\ \bibinfo {pages} {2464} (\bibinfo {year} {2019})}\BibitemShut {NoStop}%
\bibitem [{\citenamefont {Liu}\ \emph {et~al.}(2019)\citenamefont {Liu}, \citenamefont {van Baren}, \citenamefont {Taniguchi}, \citenamefont {Watanabe}, \citenamefont {Chang},\ and\ \citenamefont {Lui}}]{Liu2019PRB}%
  \BibitemOpen
  \bibfield  {author} {\bibinfo {author} {\bibfnamefont {E.}~\bibnamefont {Liu}}, \bibinfo {author} {\bibfnamefont {J.}~\bibnamefont {van Baren}}, \bibinfo {author} {\bibfnamefont {T.}~\bibnamefont {Taniguchi}}, \bibinfo {author} {\bibfnamefont {K.}~\bibnamefont {Watanabe}}, \bibinfo {author} {\bibfnamefont {Y.-C.}\ \bibnamefont {Chang}},\ and\ \bibinfo {author} {\bibfnamefont {C.~H.}\ \bibnamefont {Lui}},\ }\bibfield  {title} {\bibinfo {title} {Magnetophotoluminescence of exciton rydberg states in monolayer $\mathrm{WS}{\mathrm{e}}_{2}$},\ }\href {https://doi.org/10.1103/PhysRevB.99.205420} {\bibfield  {journal} {\bibinfo  {journal} {Phys. Rev. B}\ }\textbf {\bibinfo {volume} {99}},\ \bibinfo {pages} {205420} (\bibinfo {year} {2019})}\BibitemShut {NoStop}%
\bibitem [{\citenamefont {He}\ \emph {et~al.}(2014)\citenamefont {He}, \citenamefont {Kumar}, \citenamefont {Zhao}, \citenamefont {Wang}, \citenamefont {Mak}, \citenamefont {Zhao},\ and\ \citenamefont {Shan}}]{He2014PRL}%
  \BibitemOpen
  \bibfield  {author} {\bibinfo {author} {\bibfnamefont {K.}~\bibnamefont {He}}, \bibinfo {author} {\bibfnamefont {N.}~\bibnamefont {Kumar}}, \bibinfo {author} {\bibfnamefont {L.}~\bibnamefont {Zhao}}, \bibinfo {author} {\bibfnamefont {Z.}~\bibnamefont {Wang}}, \bibinfo {author} {\bibfnamefont {K.~F.}\ \bibnamefont {Mak}}, \bibinfo {author} {\bibfnamefont {H.}~\bibnamefont {Zhao}},\ and\ \bibinfo {author} {\bibfnamefont {J.}~\bibnamefont {Shan}},\ }\bibfield  {title} {\bibinfo {title} {{Tightly Bound Excitons in Monolayer ${\mathrm{WSe}}_{2}$}},\ }\href {https://doi.org/10.1103/PhysRevLett.113.026803} {\bibfield  {journal} {\bibinfo  {journal} {Phys. Rev. Lett.}\ }\textbf {\bibinfo {volume} {113}},\ \bibinfo {pages} {026803} (\bibinfo {year} {2014})}\BibitemShut {NoStop}%
\bibitem [{\citenamefont {Stier}\ \emph {et~al.}(2018)\citenamefont {Stier}, \citenamefont {Wilson}, \citenamefont {Velizhanin}, \citenamefont {Kono}, \citenamefont {Xu},\ and\ \citenamefont {Crooker}}]{Stier2018PRL}%
  \BibitemOpen
  \bibfield  {author} {\bibinfo {author} {\bibfnamefont {A.~V.}\ \bibnamefont {Stier}}, \bibinfo {author} {\bibfnamefont {N.~P.}\ \bibnamefont {Wilson}}, \bibinfo {author} {\bibfnamefont {K.~A.}\ \bibnamefont {Velizhanin}}, \bibinfo {author} {\bibfnamefont {J.}~\bibnamefont {Kono}}, \bibinfo {author} {\bibfnamefont {X.}~\bibnamefont {Xu}},\ and\ \bibinfo {author} {\bibfnamefont {S.~A.}\ \bibnamefont {Crooker}},\ }\bibfield  {title} {\bibinfo {title} {Magnetooptics of exciton rydberg states in a monolayer semiconductor},\ }\href {https://doi.org/10.1103/PhysRevLett.120.057405} {\bibfield  {journal} {\bibinfo  {journal} {Phys. Rev. Lett.}\ }\textbf {\bibinfo {volume} {120}},\ \bibinfo {pages} {057405} (\bibinfo {year} {2018})}\BibitemShut {NoStop}%
\bibitem [{\citenamefont {Liu}\ \emph {et~al.}(2021)\citenamefont {Liu}, \citenamefont {van Baren}, \citenamefont {Lu}, \citenamefont {Taniguchi}, \citenamefont {Watanabe}, \citenamefont {Smirnov}, \citenamefont {Chang},\ and\ \citenamefont {Lui}}]{Liu2021NatCommun}%
  \BibitemOpen
  \bibfield  {author} {\bibinfo {author} {\bibfnamefont {E.}~\bibnamefont {Liu}}, \bibinfo {author} {\bibfnamefont {J.}~\bibnamefont {van Baren}}, \bibinfo {author} {\bibfnamefont {Z.}~\bibnamefont {Lu}}, \bibinfo {author} {\bibfnamefont {T.}~\bibnamefont {Taniguchi}}, \bibinfo {author} {\bibfnamefont {K.}~\bibnamefont {Watanabe}}, \bibinfo {author} {\bibfnamefont {D.}~\bibnamefont {Smirnov}}, \bibinfo {author} {\bibfnamefont {Y.-C.}\ \bibnamefont {Chang}},\ and\ \bibinfo {author} {\bibfnamefont {C.~H.}\ \bibnamefont {Lui}},\ }\bibfield  {title} {\bibinfo {title} {{Exciton-polaron Rydberg states in monolayer MoSe$_2$ and WSe$_2$}},\ }\href {https://doi.org/10.1038/s41467-021-26304-w} {\bibfield  {journal} {\bibinfo  {journal} {Nature Communications}\ }\textbf {\bibinfo {volume} {12}},\ \bibinfo {pages} {6131} (\bibinfo {year} {2021})}\BibitemShut {NoStop}%
\bibitem [{\citenamefont {Wang}\ \emph {et~al.}(2020)\citenamefont {Wang}, \citenamefont {Li}, \citenamefont {Li}, \citenamefont {Lu}, \citenamefont {Miao}, \citenamefont {Lian}, \citenamefont {Meng}, \citenamefont {Blei}, \citenamefont {Taniguchi}, \citenamefont {Watanabe}, \citenamefont {Tongay}, \citenamefont {Smirnov}, \citenamefont {Zhang},\ and\ \citenamefont {Shi}}]{Wang2020NanoLett}%
  \BibitemOpen
  \bibfield  {author} {\bibinfo {author} {\bibfnamefont {T.}~\bibnamefont {Wang}}, \bibinfo {author} {\bibfnamefont {Z.}~\bibnamefont {Li}}, \bibinfo {author} {\bibfnamefont {Y.}~\bibnamefont {Li}}, \bibinfo {author} {\bibfnamefont {Z.}~\bibnamefont {Lu}}, \bibinfo {author} {\bibfnamefont {S.}~\bibnamefont {Miao}}, \bibinfo {author} {\bibfnamefont {Z.}~\bibnamefont {Lian}}, \bibinfo {author} {\bibfnamefont {Y.}~\bibnamefont {Meng}}, \bibinfo {author} {\bibfnamefont {M.}~\bibnamefont {Blei}}, \bibinfo {author} {\bibfnamefont {T.}~\bibnamefont {Taniguchi}}, \bibinfo {author} {\bibfnamefont {K.}~\bibnamefont {Watanabe}}, \bibinfo {author} {\bibfnamefont {S.}~\bibnamefont {Tongay}}, \bibinfo {author} {\bibfnamefont {D.}~\bibnamefont {Smirnov}}, \bibinfo {author} {\bibfnamefont {C.}~\bibnamefont {Zhang}},\ and\ \bibinfo {author} {\bibfnamefont {S.-F.}\ \bibnamefont {Shi}},\ }\bibfield  {title} {\bibinfo {title} {{Giant Valley-Polarized Rydberg Excitons in Monolayer WSe$_2$ Revealed by Magneto-photocurrent
  Spectroscopy}},\ }\href {https://doi.org/10.1021/acs.nanolett.0c03167} {\bibfield  {journal} {\bibinfo  {journal} {Nano Letters}\ }\textbf {\bibinfo {volume} {20}},\ \bibinfo {pages} {7635} (\bibinfo {year} {2020})}\BibitemShut {NoStop}%
\bibitem [{\citenamefont {Woo}\ \emph {et~al.}(2023)\citenamefont {Woo}, \citenamefont {Zobelli}, \citenamefont {Schneider}, \citenamefont {Arora}, \citenamefont {Preu\ss{}}, \citenamefont {Carey}, \citenamefont {Michaelis~de Vasconcellos}, \citenamefont {Palummo}, \citenamefont {Bratschitsch},\ and\ \citenamefont {Tizei}}]{Woo2023PRB}%
  \BibitemOpen
  \bibfield  {author} {\bibinfo {author} {\bibfnamefont {S.~Y.}\ \bibnamefont {Woo}}, \bibinfo {author} {\bibfnamefont {A.}~\bibnamefont {Zobelli}}, \bibinfo {author} {\bibfnamefont {R.}~\bibnamefont {Schneider}}, \bibinfo {author} {\bibfnamefont {A.}~\bibnamefont {Arora}}, \bibinfo {author} {\bibfnamefont {J.~A.}\ \bibnamefont {Preu\ss{}}}, \bibinfo {author} {\bibfnamefont {B.~J.}\ \bibnamefont {Carey}}, \bibinfo {author} {\bibfnamefont {S.}~\bibnamefont {Michaelis~de Vasconcellos}}, \bibinfo {author} {\bibfnamefont {M.}~\bibnamefont {Palummo}}, \bibinfo {author} {\bibfnamefont {R.}~\bibnamefont {Bratschitsch}},\ and\ \bibinfo {author} {\bibfnamefont {L.~H.~G.}\ \bibnamefont {Tizei}},\ }\bibfield  {title} {\bibinfo {title} {{Excitonic absorption signatures of twisted bilayer WSe$_2$ by electron energy-loss spectroscopy}},\ }\href {https://doi.org/10.1103/PhysRevB.107.155429} {\bibfield  {journal} {\bibinfo  {journal} {Phys. Rev. B}\ }\textbf {\bibinfo {volume} {107}},\ \bibinfo {pages} {155429} (\bibinfo
  {year} {2023})}\BibitemShut {NoStop}%
\bibitem [{\citenamefont {Arora}\ \emph {et~al.}(2015)\citenamefont {Arora}, \citenamefont {Koperski}, \citenamefont {Nogajewski}, \citenamefont {Marcus}, \citenamefont {Faugeras},\ and\ \citenamefont {Potemski}}]{Arora2015Nanoscale}%
  \BibitemOpen
  \bibfield  {author} {\bibinfo {author} {\bibfnamefont {A.}~\bibnamefont {Arora}}, \bibinfo {author} {\bibfnamefont {M.}~\bibnamefont {Koperski}}, \bibinfo {author} {\bibfnamefont {K.}~\bibnamefont {Nogajewski}}, \bibinfo {author} {\bibfnamefont {J.}~\bibnamefont {Marcus}}, \bibinfo {author} {\bibfnamefont {C.}~\bibnamefont {Faugeras}},\ and\ \bibinfo {author} {\bibfnamefont {M.}~\bibnamefont {Potemski}},\ }\bibfield  {title} {\bibinfo {title} {{Excitonic resonances in thin films of WSe$_2$: from monolayer to bulk material}},\ }\href {https://doi.org/10.1039/C5NR01536G} {\bibfield  {journal} {\bibinfo  {journal} {Nanoscale}\ }\textbf {\bibinfo {volume} {7}},\ \bibinfo {pages} {10421} (\bibinfo {year} {2015})}\BibitemShut {NoStop}%
\bibitem [{\citenamefont {Schmidt}\ \emph {et~al.}(2016)\citenamefont {Schmidt}, \citenamefont {Niehues}, \citenamefont {Schneider}, \citenamefont {Drüppel}, \citenamefont {Deilmann}, \citenamefont {Rohlfing}, \citenamefont {de~Vasconcellos}, \citenamefont {Castellanos-Gomez},\ and\ \citenamefont {Bratschitsch}}]{Schmidt20162DMater}%
  \BibitemOpen
  \bibfield  {author} {\bibinfo {author} {\bibfnamefont {R.}~\bibnamefont {Schmidt}}, \bibinfo {author} {\bibfnamefont {I.}~\bibnamefont {Niehues}}, \bibinfo {author} {\bibfnamefont {R.}~\bibnamefont {Schneider}}, \bibinfo {author} {\bibfnamefont {M.}~\bibnamefont {Drüppel}}, \bibinfo {author} {\bibfnamefont {T.}~\bibnamefont {Deilmann}}, \bibinfo {author} {\bibfnamefont {M.}~\bibnamefont {Rohlfing}}, \bibinfo {author} {\bibfnamefont {S.~M.}\ \bibnamefont {de~Vasconcellos}}, \bibinfo {author} {\bibfnamefont {A.}~\bibnamefont {Castellanos-Gomez}},\ and\ \bibinfo {author} {\bibfnamefont {R.}~\bibnamefont {Bratschitsch}},\ }\bibfield  {title} {\bibinfo {title} {{Reversible uniaxial strain tuning in atomically thin WSe$_2$}},\ }\href {https://doi.org/10.1088/2053-1583/3/2/021011} {\bibfield  {journal} {\bibinfo  {journal} {2D Materials}\ }\textbf {\bibinfo {volume} {3}},\ \bibinfo {pages} {021011} (\bibinfo {year} {2016})}\BibitemShut {NoStop}%
\bibitem [{\citenamefont {Hong}\ \emph {et~al.}(2020)\citenamefont {Hong}, \citenamefont {Senga}, \citenamefont {Pichler},\ and\ \citenamefont {Suenaga}}]{Hong2020PRL}%
  \BibitemOpen
  \bibfield  {author} {\bibinfo {author} {\bibfnamefont {J.}~\bibnamefont {Hong}}, \bibinfo {author} {\bibfnamefont {R.}~\bibnamefont {Senga}}, \bibinfo {author} {\bibfnamefont {T.}~\bibnamefont {Pichler}},\ and\ \bibinfo {author} {\bibfnamefont {K.}~\bibnamefont {Suenaga}},\ }\bibfield  {title} {\bibinfo {title} {{Probing Exciton Dispersions of Freestanding Monolayer ${\mathrm{WSe}}_{2}$ by Momentum-Resolved Electron Energy-Loss Spectroscopy}},\ }\href {https://doi.org/10.1103/PhysRevLett.124.087401} {\bibfield  {journal} {\bibinfo  {journal} {Phys. Rev. Lett.}\ }\textbf {\bibinfo {volume} {124}},\ \bibinfo {pages} {087401} (\bibinfo {year} {2020})}\BibitemShut {NoStop}%
\bibitem [{\citenamefont {Wang}\ \emph {et~al.}(2015)\citenamefont {Wang}, \citenamefont {Marie}, \citenamefont {Gerber}, \citenamefont {Amand}, \citenamefont {Lagarde}, \citenamefont {Bouet}, \citenamefont {Vidal}, \citenamefont {Balocchi},\ and\ \citenamefont {Urbaszek}}]{Wang2015PRL}%
  \BibitemOpen
  \bibfield  {author} {\bibinfo {author} {\bibfnamefont {G.}~\bibnamefont {Wang}}, \bibinfo {author} {\bibfnamefont {X.}~\bibnamefont {Marie}}, \bibinfo {author} {\bibfnamefont {I.}~\bibnamefont {Gerber}}, \bibinfo {author} {\bibfnamefont {T.}~\bibnamefont {Amand}}, \bibinfo {author} {\bibfnamefont {D.}~\bibnamefont {Lagarde}}, \bibinfo {author} {\bibfnamefont {L.}~\bibnamefont {Bouet}}, \bibinfo {author} {\bibfnamefont {M.}~\bibnamefont {Vidal}}, \bibinfo {author} {\bibfnamefont {A.}~\bibnamefont {Balocchi}},\ and\ \bibinfo {author} {\bibfnamefont {B.}~\bibnamefont {Urbaszek}},\ }\bibfield  {title} {\bibinfo {title} {{Giant Enhancement of the Optical Second-Harmonic Emission of ${\mathrm{WSe}}_{2}$ Monolayers by Laser Excitation at Exciton Resonances}},\ }\href {https://doi.org/10.1103/PhysRevLett.114.097403} {\bibfield  {journal} {\bibinfo  {journal} {Phys. Rev. Lett.}\ }\textbf {\bibinfo {volume} {114}},\ \bibinfo {pages} {097403} (\bibinfo {year} {2015})}\BibitemShut {NoStop}%
\bibitem [{\citenamefont {Li}\ \emph {et~al.}(2014)\citenamefont {Li}, \citenamefont {Chernikov}, \citenamefont {Zhang}, \citenamefont {Rigosi}, \citenamefont {Hill}, \citenamefont {van~der Zande}, \citenamefont {Chenet}, \citenamefont {Shih}, \citenamefont {Hone},\ and\ \citenamefont {Heinz}}]{Li2014PRB}%
  \BibitemOpen
  \bibfield  {author} {\bibinfo {author} {\bibfnamefont {Y.}~\bibnamefont {Li}}, \bibinfo {author} {\bibfnamefont {A.}~\bibnamefont {Chernikov}}, \bibinfo {author} {\bibfnamefont {X.}~\bibnamefont {Zhang}}, \bibinfo {author} {\bibfnamefont {A.}~\bibnamefont {Rigosi}}, \bibinfo {author} {\bibfnamefont {H.~M.}\ \bibnamefont {Hill}}, \bibinfo {author} {\bibfnamefont {A.~M.}\ \bibnamefont {van~der Zande}}, \bibinfo {author} {\bibfnamefont {D.~A.}\ \bibnamefont {Chenet}}, \bibinfo {author} {\bibfnamefont {E.-M.}\ \bibnamefont {Shih}}, \bibinfo {author} {\bibfnamefont {J.}~\bibnamefont {Hone}},\ and\ \bibinfo {author} {\bibfnamefont {T.~F.}\ \bibnamefont {Heinz}},\ }\bibfield  {title} {\bibinfo {title} {{Measurement of the optical dielectric function of monolayer transition-metal dichalcogenides: ${\mathrm{MoS}}_{2}$, $\mathrm{Mo}\mathrm{S}{\mathrm{e}}_{2}$, ${\mathrm{WS}}_{2}$, and $\mathrm{WS}{\mathrm{e}}_{2}$}},\ }\href {https://doi.org/10.1103/PhysRevB.90.205422} {\bibfield  {journal} {\bibinfo  {journal}
  {Phys. Rev. B}\ }\textbf {\bibinfo {volume} {90}},\ \bibinfo {pages} {205422} (\bibinfo {year} {2014})}\BibitemShut {NoStop}%
\bibitem [{\citenamefont {Wierzbowski}\ \emph {et~al.}(2017)\citenamefont {Wierzbowski}, \citenamefont {Klein}, \citenamefont {Sigger}, \citenamefont {Straubinger}, \citenamefont {Kremser}, \citenamefont {Taniguchi}, \citenamefont {Watanabe}, \citenamefont {Wurstbauer}, \citenamefont {Holleitner}, \citenamefont {Kaniber}, \citenamefont {M{\"u}ller},\ and\ \citenamefont {Finley}}]{Wierzbowski2017SciRep}%
  \BibitemOpen
  \bibfield  {author} {\bibinfo {author} {\bibfnamefont {J.}~\bibnamefont {Wierzbowski}}, \bibinfo {author} {\bibfnamefont {J.}~\bibnamefont {Klein}}, \bibinfo {author} {\bibfnamefont {F.}~\bibnamefont {Sigger}}, \bibinfo {author} {\bibfnamefont {C.}~\bibnamefont {Straubinger}}, \bibinfo {author} {\bibfnamefont {M.}~\bibnamefont {Kremser}}, \bibinfo {author} {\bibfnamefont {T.}~\bibnamefont {Taniguchi}}, \bibinfo {author} {\bibfnamefont {K.}~\bibnamefont {Watanabe}}, \bibinfo {author} {\bibfnamefont {U.}~\bibnamefont {Wurstbauer}}, \bibinfo {author} {\bibfnamefont {A.~W.}\ \bibnamefont {Holleitner}}, \bibinfo {author} {\bibfnamefont {M.}~\bibnamefont {Kaniber}}, \bibinfo {author} {\bibfnamefont {K.}~\bibnamefont {M{\"u}ller}},\ and\ \bibinfo {author} {\bibfnamefont {J.~J.}\ \bibnamefont {Finley}},\ }\bibfield  {title} {\bibinfo {title} {Direct exciton emission from atomically thin transition metal dichalcogenide heterostructures near the lifetime limit},\ }\href {https://doi.org/10.1038/s41598-017-09739-4}
  {\bibfield  {journal} {\bibinfo  {journal} {Scientific Reports}\ }\textbf {\bibinfo {volume} {7}},\ \bibinfo {pages} {12383} (\bibinfo {year} {2017})}\BibitemShut {NoStop}%
\bibitem [{\citenamefont {Madéo}\ \emph {et~al.}(2020)\citenamefont {Madéo}, \citenamefont {Man}, \citenamefont {Sahoo}, \citenamefont {Campbell}, \citenamefont {Pareek}, \citenamefont {Wong}, \citenamefont {Al-Mahboob}, \citenamefont {Chan}, \citenamefont {Karmakar}, \citenamefont {Mariserla}, \citenamefont {Li}, \citenamefont {Heinz}, \citenamefont {Cao},\ and\ \citenamefont {Dani}}]{Madeo2020Sicence}%
  \BibitemOpen
  \bibfield  {author} {\bibinfo {author} {\bibfnamefont {J.}~\bibnamefont {Madéo}}, \bibinfo {author} {\bibfnamefont {M.~K.~L.}\ \bibnamefont {Man}}, \bibinfo {author} {\bibfnamefont {C.}~\bibnamefont {Sahoo}}, \bibinfo {author} {\bibfnamefont {M.}~\bibnamefont {Campbell}}, \bibinfo {author} {\bibfnamefont {V.}~\bibnamefont {Pareek}}, \bibinfo {author} {\bibfnamefont {E.~L.}\ \bibnamefont {Wong}}, \bibinfo {author} {\bibfnamefont {A.}~\bibnamefont {Al-Mahboob}}, \bibinfo {author} {\bibfnamefont {N.~S.}\ \bibnamefont {Chan}}, \bibinfo {author} {\bibfnamefont {A.}~\bibnamefont {Karmakar}}, \bibinfo {author} {\bibfnamefont {B.~M.~K.}\ \bibnamefont {Mariserla}}, \bibinfo {author} {\bibfnamefont {X.}~\bibnamefont {Li}}, \bibinfo {author} {\bibfnamefont {T.~F.}\ \bibnamefont {Heinz}}, \bibinfo {author} {\bibfnamefont {T.}~\bibnamefont {Cao}},\ and\ \bibinfo {author} {\bibfnamefont {K.~M.}\ \bibnamefont {Dani}},\ }\bibfield  {title} {\bibinfo {title} {Directly visualizing the momentum-forbidden dark excitons and
  their dynamics in atomically thin semiconductors},\ }\href {https://doi.org/10.1126/science.aba1029} {\bibfield  {journal} {\bibinfo  {journal} {Science}\ }\textbf {\bibinfo {volume} {370}},\ \bibinfo {pages} {1199} (\bibinfo {year} {2020})}\BibitemShut {NoStop}%
\bibitem [{\citenamefont {Li}\ \emph {et~al.}(2020)\citenamefont {Li}, \citenamefont {Goryca}, \citenamefont {Wilson}, \citenamefont {Stier}, \citenamefont {Xu},\ and\ \citenamefont {Crooker}}]{Li2020PRL}%
  \BibitemOpen
  \bibfield  {author} {\bibinfo {author} {\bibfnamefont {J.}~\bibnamefont {Li}}, \bibinfo {author} {\bibfnamefont {M.}~\bibnamefont {Goryca}}, \bibinfo {author} {\bibfnamefont {N.~P.}\ \bibnamefont {Wilson}}, \bibinfo {author} {\bibfnamefont {A.~V.}\ \bibnamefont {Stier}}, \bibinfo {author} {\bibfnamefont {X.}~\bibnamefont {Xu}},\ and\ \bibinfo {author} {\bibfnamefont {S.~A.}\ \bibnamefont {Crooker}},\ }\bibfield  {title} {\bibinfo {title} {Spontaneous valley polarization of interacting carriers in a monolayer semiconductor},\ }\href {https://doi.org/10.1103/PhysRevLett.125.147602} {\bibfield  {journal} {\bibinfo  {journal} {Phys. Rev. Lett.}\ }\textbf {\bibinfo {volume} {125}},\ \bibinfo {pages} {147602} (\bibinfo {year} {2020})}\BibitemShut {NoStop}%
\bibitem [{\citenamefont {Deilmann}\ and\ \citenamefont {Thygesen}(2017)}]{Deilmann2017PRB}%
  \BibitemOpen
  \bibfield  {author} {\bibinfo {author} {\bibfnamefont {T.}~\bibnamefont {Deilmann}}\ and\ \bibinfo {author} {\bibfnamefont {K.~S.}\ \bibnamefont {Thygesen}},\ }\bibfield  {title} {\bibinfo {title} {Dark excitations in monolayer transition metal dichalcogenides},\ }\href {https://doi.org/10.1103/PhysRevB.96.201113} {\bibfield  {journal} {\bibinfo  {journal} {Phys. Rev. B}\ }\textbf {\bibinfo {volume} {96}},\ \bibinfo {pages} {201113} (\bibinfo {year} {2017})}\BibitemShut {NoStop}%
\bibitem [{\citenamefont {Deilmann}\ and\ \citenamefont {Thygesen}(2019)}]{Deilmann20192DMater}%
  \BibitemOpen
  \bibfield  {author} {\bibinfo {author} {\bibfnamefont {T.}~\bibnamefont {Deilmann}}\ and\ \bibinfo {author} {\bibfnamefont {K.~S.}\ \bibnamefont {Thygesen}},\ }\bibfield  {title} {\bibinfo {title} {Finite-momentum exciton landscape in mono- and bilayer transition metal dichalcogenides},\ }\href {https://doi.org/10.1088/2053-1583/ab0e1d} {\bibfield  {journal} {\bibinfo  {journal} {2D Materials}\ }\textbf {\bibinfo {volume} {6}},\ \bibinfo {pages} {035003} (\bibinfo {year} {2019})}\BibitemShut {NoStop}%
\bibitem [{\citenamefont {Song}\ and\ \citenamefont {Dery}(2013)}]{Song2013}%
  \BibitemOpen
  \bibfield  {author} {\bibinfo {author} {\bibfnamefont {Y.}~\bibnamefont {Song}}\ and\ \bibinfo {author} {\bibfnamefont {H.}~\bibnamefont {Dery}},\ }\bibfield  {title} {\bibinfo {title} {{Transport Theory of Monolayer Transition-Metal Dichalcogenides through Symmetry}},\ }\href {https://doi.org/10.1103/PhysRevLett.111.026601} {\bibfield  {journal} {\bibinfo  {journal} {Phys. Rev. Lett.}\ }\textbf {\bibinfo {volume} {111}},\ \bibinfo {pages} {026601} (\bibinfo {year} {2013})}\BibitemShut {NoStop}%
\bibitem [{\citenamefont {Atabek}\ \emph {et~al.}(1974)\citenamefont {Atabek}, \citenamefont {Deutsch},\ and\ \citenamefont {Lavaud}}]{Atabek74}%
  \BibitemOpen
  \bibfield  {author} {\bibinfo {author} {\bibfnamefont {O.}~\bibnamefont {Atabek}}, \bibinfo {author} {\bibfnamefont {C.}~\bibnamefont {Deutsch}},\ and\ \bibinfo {author} {\bibfnamefont {M.}~\bibnamefont {Lavaud}},\ }\bibfield  {title} {\bibinfo {title} {{Schr\"odinger equation for the two-dimensional Coulomb potential}},\ }\href {https://doi.org/10.1103/PhysRevA.9.2617} {\bibfield  {journal} {\bibinfo  {journal} {Phys. Rev. A}\ }\textbf {\bibinfo {volume} {9}},\ \bibinfo {pages} {2617} (\bibinfo {year} {1974})}\BibitemShut {NoStop}%
\end{thebibliography}%
\end{document}